\documentclass[12pt]{article}

\usepackage[font=small,labelfont=bf]{caption}
\usepackage{subcaption}

\usepackage[margin=1in]{geometry}
\usepackage{amssymb}
\usepackage{amsmath}
\usepackage{dsfont}
\usepackage{amsfonts}
\usepackage{bm}
\usepackage{xfrac}
\usepackage{subcaption}
\usepackage[hidelinks]{hyperref}
\usepackage{authblk}
\usepackage[square,numbers]{natbib}
\usepackage{rotating}
\usepackage{booktabs}

\newtheorem{definition}{Definition}

\DeclareMathOperator*{\argmax}{arg\,max}

\newcommand{\qunderline}[1]{\hbox{\oalign{$\bm{#1}$\crcr\hidewidth$\scriptscriptstyle\bm{\sim}$\hidewidth}}}

\newcommand{\p}{\mathds{P}} 
\newcommand{\tb}[1]{\text{\textbf{#1}}} 
\newcommand{\ith}{$i$-th~} 
\newcommand{\jth}{$j$-th~}
\newcommand{\kth}{$k$-th~}
\newcommand{\M}{\mathcal{M}}

\title{An Exposure Model Framework for Signal Detection based on Electronic Healthcare Data}

\author[a]{Louis Dijkstra}
\author[a]{Tania Schink}
\author[a]{Ronja Foraita\footnote{Corresponding author. E-mail: \url{foraita@leibniz-bips.de}}} 

\affil[a]{Leibniz Institute for Prevention Research \& Epidemiology -- BIPS, \newline Achterstra{\ss}e 30, 28359 Bremen, Germany}

\begin{document}

\maketitle

\begin{abstract}
\noindent Despite extensive safety assessments of drugs prior to their introduction to the market, certain adverse drug reactions (ADRs) remain undetected. The primary objective of pharmacovigilance is to identify these ADRs (i.e., signals).
In addition to traditional spontaneous reporting systems (SRSs), electronic health (EHC) data is being used for signal detection as well. Unlike SRS, EHC data is longitudinal and thus requires assumptions about the patient's drug exposure history and its impact on ADR occurrences over time, which many current methods do implicitly.

We propose an exposure model framework that explicitly models the longitudinal relationship between the drug and the ADR. By considering multiple such models simultaneously, we can detect signals that might be missed by other approaches. The parameters of these models are estimated using maximum likelihood, and the Bayesian Information Criterion (BIC) is employed to select the most suitable model. Since BIC is connected to the posterior distribution, it servers the dual purpose of identifying the best-fitting model and determining the presence of a signal by evaluating the posterior probability of the null model.

We evaluate the effectiveness of this framework through a simulation study, for which we develop an EHC data simulator. Additionally, we conduct a case study applying our approach to four drug-ADR pairs using an EHC dataset comprising over 1.2 million insured individuals. Both the method and the EHC data simulator code are publicly accessible as part of the \texttt{R} package \texttt{https://github.com/bips-hb/expard}.
\begin{flushleft}
\end{flushleft}
\end{abstract}


\renewcommand\thefootnote{\fnsymbol{footnote}}
\setcounter{footnote}{1}

\section{Introduction}\label{sec:introduction}
Despite the thorough examination of drugs for potential side effects before their market release, 
some adverse drug reactions (ADRs) may go unnoticed until after the drug enters the market \citep{WHO2002,suling2012,harpaz2012,bailey2016,waller2017}. This could be due to several reasons. 
The pivotal randomized clinical trials (RCTs) are designed with a focus on assessing efficacy, resulting in sample sizes often inadequate for detecting rare safety outcomes \citep{grosso2011,singh2012}. In these trials, patients typically need to meet multiple inclusion criteria, and vulnerable groups, such as pregnant women, the elderly \citep{budnitz2011}, and individuals with multiple health conditions, are frequently either excluded or underrepresented \citep{mosenifar2007,feldman2019}. 
Furthermore, patients in RCTs are monitored for a limited duration only, making it difficult to uncover any potential long-term effects. 
For that reason, pharmacovigilance \citep{WHO2002,waller2017,finney1965} plays a pivotal role in ensuring the safety of pharmaceutical products. 

With the aim of promptly identifying drugs that may pose health risks, spontaneous reporting systems (SRSs) have been established over the years \citep{routledge1998}. Healthcare professionals, pharmaceutical companies, and, in some cases, patients \citep{margraff2014} can submit a spontaneous report to such a system when they suspect a drug may be associated with a previously unknown reaction \citep{suling2012,bailey2016,waller2017,schroeder1998}. These reports are collected, cleaned, stored, and subsequently analyzed by a committee of medical experts \citep{arnaiz2001,onakpoya2016, alomar2019}. Due to the sheer volume of accumulated reports typically contained in these systems \citep{vigibase}, the idea was to present the committee of medical experts with an automatically curated list of drug-ADR pairs that needed their attention, rather than the raw reports themselves \citep{tubert1991,tubert1992,agrawal1993}. 
Each entry on such a list is referred to as a a \emph{signal}. Therefore, the process of creating such a shortlist is also known as \emph{signal detection} \citep{waller2017}.

Even though SRSs form the cornerstone of pharmacovigilance, they also come with several limitations. First, the total number of patients exposed to the drug or experiencing the ADR is essentially unknown. Reports are submitted only when a patient has both been exposed to the drug \emph{and} has experienced the ADR, leading to what is known as the unknown denominator problem \citep{dumouchel1999}. Secondly, there is  potentially an over- or underreporting bias \citep{schroeder1998,tubert1992,moride1997,vanderheijden2002,hazell2006}. Newly introduced drugs, for example, may attract more attention from healthcare professionals and are, thus, more likely to be reported. Third, the decision on what information to include in the report is made on an individual basis. For instance, one might choose to exclude a drug if it is deemed too unlikely to have caused the ADR or if the patient's exposure is considered too far in the past.

Over the last two decades, alongside SRS data,  pharmacovigilance has begun considering longitudinal data as well, specifically \emph{electronic healthcare} (EHC) \emph{data} \citep{coste2022,shin2022,dijkstra2020,karlsson2013,murray2011,schuemie2011,omop2010}.
EHC data include individual patients' drug prescriptions and medical events over time, along with personal details such as age, sex, and place of residence \citep{suling2012, coloma2011, zorych2013}. This kind of data, to some extent, mitigates the limitations of SRS data mentioned earlier \citep{reps2013}. The total number of patients exposed to the drug \emph{and/or} experiencing the ADR is known; there is a reduction in over- and underreporting bias \citep{harpaz2010}, and it eliminates the need to make choices about what to report. An additional advantage is that EHC data sets can be very extensive, containing up to millions of patients \citep{haug2021}. 


Another distinction between EHC and SRS data lies in their timeliness \citep{tubert-bitter2016}. Reports are typically submitted promptly to SRSs, whereas EHC data often experience delays. Consequently, SRS data play a crucial role in detecting ADRs of drugs recently introduced to the market. On the other hand, the richness of EHC data may prove beneficial in identifying more subtle associations between the drug and ADRs, a task that might be challenging on the basis of SRS data alone \citep{trifiro2011}.

A plethora of signal detection methods have been proposed for EHC data \citep{suling2012,schuemie2011,coloma2011,zorych2013,vangaalen2015,vangaalen2017,kelly2023}. To address its longitudinal nature, many of these methods convert EHC data into a format resembling SRS data and utilize techniques initially developed for SRSs \citep{zorych2013}. Other approaches such as LASSO \citep{dijkstra2022discovery} and Random Forests (RFs) condense a patient's exposures and ADR occurrences over time into a limited number of variables while striving to retain some of its temporal information \citep{dijkstra2022discovery}.

By transforming EHC data in this way, one implicitly makes assumptions about the temporal relationship between drug exposure and the ADR \citep{noren2013}. 
Van Gaalen et al. \citep{vangaalen2015,vangaalen2017} introduce the concept of an \emph{exposure model}, i.e., a formal description of how exposure to the drug affects the risk for a patient to experience the ADR over time. The idea is to define multiple exposure models, each attempting to capture a different type of temporal relationship. For instance, one can establish an exposure model for withdrawal effects, where the risk of experiencing the ADR peaks just after the patient's exposure stops and diminishes rather quickly. Alternatively, a model can be defined where the effect of the exposure is long-term, and the risk increases gradually over time \citep{vangaalen2015}. By considering multiple of these models simultaneously for each drug-ADR pair, the hope is to be able to identify many different temporal relationships, some of which may not be discernible using conventional approaches since the effect is diluted. Van Gaalen et al. \citep{vangaalen2015,vangaalen2017} define several such exposure models. However, the range of models is limited and it is unclear how the parameters of these models are to be estimated on the basis of the data. Additionally, their approach focuses on a single drug-ADR pair, leaving ambiguity about its applicability in a pharmacovigilance context where multiple drug-ADR pairs are considered simultaneously.

An alternative approach involves employing cubic B-splines to characterize the connection between the drug-ADR pair \citep{kelly2023}. While this approach offers flexibility, it comes with drawbacks, as it does not explicitly define the nature of the relationship between the drug and ADR pair. More critically, the application of cubic B-splines in the traditional pharmacovigilance setting, where multiple drug-ADR pairs are simultaneously considered, is unclear as well \citep{kelly2023}. 

In this work, we present a comprehensive exposure model framework for pharmacovigilance based on EHC data to address some of the aforementioned limitations. We formally define the concept of an exposure model. In addition, we propose  eight exposure models, known to occur, at least approximately, in real-world data. It is important to note that these serve as examples, and any other exposure model can be defined within the framework. The exposure models' parameters are then estimated using maximum likelihood.

We advocate the use of the Bayesian Information Criterion (BIC) for model selection, given its focus on both the quality of data fit (expressed by the model's likelihood) and the model's complexity (quantified in terms of the number of parameters). An additional advantage of the BIC is its close relationship to the posterior probability of the exposure model. By utilizing the posterior probability of the null model, indicating no association between the drug and the ADR (see  Section~\ref{sec:modelSelection}), our approach can effectively achieve two objectives: 1) determining the presence of an association, and 2) identifying which exposure model among those considered best fits the data.

The posterior probabilities of the exposure models offer an added advantage, making the approach well-suited for pharmacovigilance by providing a basis for which drug-ADR pairs are to be considered a signal. 
The weakest signal shows the pair with the highest probability for the null model indicating no association, while the pair with the lowest posterior probability is considered the strongest signal.

To assess the effectiveness of our approach, we perform a simulation study. We offer a simulator for EHC data, which enables users to utilize the exposure model of their choice to simulate the occurrences of ADRs over time. We then assess the performance of the method in terms of its capability to both detect an association and accurately identify the correct exposure model.

Furthermore, we demonstrate the applicability of our approach through a case study based on data from the German Pharmacoepidemiological Research Database\citep{haug2021} (GePaRD), see Section~\ref{sec:caseStudy}. In this case study, we examine four drug-ADR pairs, three of which are known to be associated, and one serving as a negative control. The literature provides information on the longitudinal nature of the associations for these three positive pairs. We assess the extent to which our method can accurately identify the correct exposure model.

The paper is structured as follows: We begin by formally defining an EHC dataset in Section~\ref{sec:definitionEHCData}. Subsequently, in Section~\ref{sec:definitionExposureModel}, we introduce a formal definition of an exposure model. Section~\ref{sec:exampleExposureModels} contains eight examples of such models. We then detail how the parameters of the exposure models can be estimated using maximum likelihood, with analytical solutions derived for three of the eight models. The remaining models are solved numerically.
Section~\ref{sec:modelSelection} considers model selection and the decision criterion for determining whether a drug is associated with an ADR.

Subsequently, we introduce the simulator for EHC data in Section~\ref{sec:simulatingEHCData} and outline the simulation set-up for the simulation study in Section~\ref{sec:simulationSetUp}. The method for assessing the performance in the simulation study is described in Section~\ref{sec:performanceAssessment}. Following this, we introduce our case study in Section~\ref{sec:caseStudy}. The outcomes of both the simulation and the case study are presented in Section~\ref{sec:results}. We conclude with some final remarks and discussion in Section~\ref{sec:conclusions}.

All the code is publicly available online. Both the simulator and the implementation of the method are available in form of the \texttt{R}~package \texttt{expard} at \url{https://github.com/bips-hb/expard}. The code related to the simulation study and case study can be found at \url{https://github.com/bips-hb/expard-simulation-study} and {\url{https://github.com/bips-hb/expard-case-study}}, respectively.

\section{Methods} \label{sec:methods}


\subsection{A Formalization of Electronic Healthcare Data} \label{sec:definitionEHCData}

Electronic healthcare data contains for multiple patients 1) the drugs they were exposed to, and 2) the ADRs they experienced over time. We denote the number of drugs on the market by $m$;  the number of registered ADRs and the number of observed patients are denoted by $n$ and $N$, respectively. The number of time points for which a patient was observed, can differ from patient to patient. We denote the total number of time points for the \kth patient by $T_k \geq 1$. We assume that each patient was observed continuously, i.e., without interruptions.  

We represent the \kth patient's drug exposure to the \ith drug over time as a random binary $T_k$-dimensional vector:  
\begin{equation*}
    \bm{X}_i^k = \left(X_i^k(1), X_i^k(2), \ldots, X_i^k(T_k) \right),
\end{equation*}
where $X_i^k(t) = 1$ if the \kth patient was exposed to the \ith drug at time point $t$, and $0$ otherwise. Likewise, the occurrences of the \jth ADR are represented by the random $T_k$-dimensional binary vector
\begin{equation*}
    \bm{Y}_j^k = \left(Y_j^k(1), Y_j^k(2), \ldots, Y_j^k(T_k) \right),
    \label{eq:definitionADRHistory}
\end{equation*}
where $Y_j^k(t) = 1$ if the \kth patient had the \jth ADR at time point $t$, and $0$ otherwise. Since there are $m$ drugs, we can represent all drug exposures for a patient $k$ as a set of $m$ different $T_k$-dimensional binary vectors, i.e., 
\begin{equation*}
    \bm{P}_{\text{drugs}}^k = \left\{\bm{X}_1^k, \bm{X}_2^k, \ldots, \bm{X}_m^k \right \}. 
    \label{eq:definitionAllDrugHistoriesPatient}
\end{equation*}
Likewise, we can represent the ADR history for patient $k$ as a set with $n$ binary vectors:  
\begin{equation*}
    \bm{P}_{\text{ADRs}}^k = \left\{\bm{Y}_1^k, \bm{Y}_2^k, \ldots, \bm{Y}_n^k \right \}. 
    \label{eq:definitionAllADRHistoriesPatient}
\end{equation*}
The \kth patient is represented by both his/her drug exposure and ADR history, i.e.,  
$$
\bm{P}_k = \left\{\bm{P}_{\text{drugs}}^k,   \bm{P}_{\text{ADRs}}^k\right\}.
$$ 
And lastly, an EHC data set is then a collection of $N$ patients: 
\begin{equation*}
    \text{\textbf{EHC}} = \left\{ \bm{P}_1 , \bm{P}_2, \ldots, \bm{P}_N \right\}. 
    \label{eq:defintionEHC}
\end{equation*}
Observations are denoted by lower-case letters: $\tb{ehc} = \{\bm{p}_k\}_{k = 1}^N$ is a given EHC data set, where $\bm{p}_k = \{\bm{p}_\text{drugs}^k, \bm{p}_\text{ADRs}^k \}$ is the \kth patient. The set $\bm{p}_\text{drugs}^k$ represents the observed drug exposures: $\bm{x}^k_i = (x_i^k(1), x_i^k(2), \ldots, x_i^k(T))$ for $i = 1,2,\ldots,m$. Similarly, the set $\bm{p}_\text{ADRs}^k$ represents the observed ADRs: $\bm{y}^k_j = (y_j^k(1), y_j^k(2), \ldots, y_j^k(T))$ for $j = 1,2,\ldots,n$.

One commonly assumes \emph{patient independence}, meaning that the joint probability density function of $\tb{EHC}$ can be factorized as 
\begin{equation*}
    \p(\tb{EHC} = \tb{ehc}) = \prod_{k = 1}^N \p(\bm{P} = \bm{p}_k), 
    \label{eq:patientIndependenceAssumption}
\end{equation*}
where $\p(\bm{P})$ denotes the probability density function of a single patient. For readability's sake, we assume in the following that all patients have been observed for the same number of time points, i.e., $T = T_1 = T_2 = \ldots = T_N$. It is straightforward to extend these signal detection methods to deal with varying observation times. 
Throughout this paper, we mainly consider single drug-ADR pairs. We, therefore, omit the subscripts $i$ and $j$ wherever possible. For ease of notation, we write 
\begin{equation*}
    \bm{X}(1 : t) = \left(X(1), X(2), \ldots, X(t - 1), X(t) \right) \in \{0,1\}^{t}
\end{equation*}
for the drug exposures for that patient from time point $1$ to $t$. 

\subsection{The Exposure Model} \label{sec:definitionExposureModel}

In pharmacovigilance, we are commonly interested in the joint probability distribution of $\bm{X}$ and $\bm{Y}$ for all drug-ADR pairs in the data set. Specifically, we are interested in whether they are independent, i.e., 
$$
    \p\left(\bm{X} = \bm{x}, \bm{Y} = \bm{y}\right) = \p\left(\bm{X} = \bm{x} \right)\p\left(\bm{Y} = \bm{y}\right)
$$
for all $\bm{x}, \bm{y} \in \{0,1\}^T$. A full specification of the joint probability density function of $(\bm{X}, \bm{Y})$ is infeasible, since it requires to specify a total of $2^{2T}$ probabilities. The model, therefore, has to be simplified. As mentioned in Section~1, signal detection methods that have been proposed in the past do this implicitly, most commonly by transforming the EHC data for a given drug-ADR pair to a single $2 \times 2$ contingency table. There are a variety of ways to do this, see, for example, the work by Zorych et al. \citep{zorych2013}. 

Instead of the full probability distribution, we consider here the conditional distribution of the ADR history given the exposures to the drug over time:
\begin{equation}
    \p\left(\bm{Y} \mid \bm{X} \right ) = \prod_{t = 1}^T \p\left(Y(t) \mid \bm{X}(1:t)\right ).
    \label{eq:conditionalDensityFunction}
\end{equation}
Modelling the full probability distribution $\p\left(\bm{X}, \bm{Y} \right ) = \p\left(\bm{Y} \mid \bm{X} \right ) \p(\bm{X})$ would require to model the drug exposure to the drug over time, i.e., $\p(\bm{X})$, which is not of direct interest. We assume that the occurrences of the ADR at different time points are independent given the drug exposure history. This can be a rather strong assumption for certain types of ADRs, e.g., anaphylaxis and myocardial infarction. We address this in the discussion, see Section~\ref{sec:conclusions}. Note that we can express $\p\left(\bm{Y}(t) \mid \bm{X}\right)$ as  $ \p\left(\bm{Y}(t) \mid \bm{X}(1:t) \right)$ since the occurrence of the \jth ADR at time point $t$ is independent of drug exposures in the future given the drug exposures up to that point in time. 

\begin{figure}
    \centering
    \includegraphics[width=.6\textwidth]{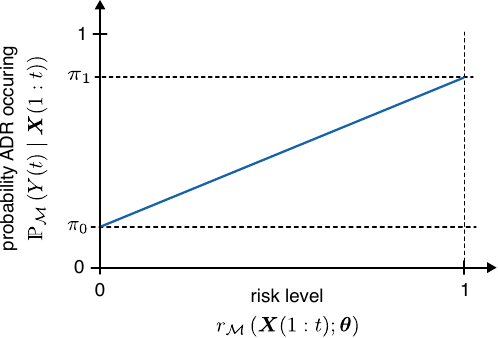}
    \caption{The probability of the ADR occurring at time point $t$, denoted as $\p_\mathcal{M}\left(Y(t) \mid \bm{X}(1:t)\right)$, plotted against the risk level represented by $r_{\mathcal{M}}\left(\bm{X}(1:t); \bm{\theta} \right)$. At the minimal risk level (0), the probability of the ADR taking place is $\pi_0$. Conversely, at the maximal risk level (1), the probability is $\pi_1$.}
\label{fig:explanationRelationshipProbabilityRisk}
\end{figure}

Here, we set out to create a framework for modelling the conditional probability distribution from eq.~\eqref{eq:conditionalDensityFunction}.  
To this end, we first define a \emph{risk level} as a value between $[0,1]$, where $0$ represents that the patient is, at that point in time, at `minimal' risk of experiencing the ADR and $1$ represents `maximal' risk (its precise meaning becomes clear later). In addition, we define $\mathcal{D}_T = \{\{0,1\}^t: t = 1,2,\ldots,T \}$ be the set of all binary vectors of length $t = 1,2,\ldots,T$. It, thus, represents all possible drug exposure histories.
We propose to represent the conditional distribution in terms of an exposure model: 
\begin{definition}
    An exposure model $\mathcal{M}$ for a drug-ADR pair is given by the tuple 
    \begin{equation}
        \left< \pi_1, \pi_0, T, r_\mathcal{M}(\cdot; \bm{\theta}), \Theta_\mathcal{M}  \right>, 
        \label{eq:definitionExposureModel}
    \end{equation}
    where $\pi_1, \pi_0 \in [0,1]$, $T$ is the number of time points, and $r_\mathcal{M}: \mathcal{D}_T \rightarrow [0,1]$ is the \textbf{risk function} that maps each binary vector in $\mathcal{D}_T$ to a risk level in the interval $[0,1]$. The risk function is parameterized by the $m$-dimensional parameter vector $\bm{\theta} = \left(\bm{\xi}; \bm{\phi}\right) \in \Theta_\mathcal{M}$, where $\bm{\xi} = \left(\xi_1, \ldots, \xi_q\right) \in \Xi_\mathcal{M}$ are continuous and $\bm{\phi} = \left(\phi_1, \ldots, \phi_s\right) \in \Phi_\mathcal{M}$ are discrete parameters $(m = q + s \geq 0)$. 
\end{definition}
Utilizing the exposure model $\mathcal{M}$, we can express the conditional probability of the ADR occurring at time point $t$, given the drug history up to that point, as
\begin{equation*}
    \p_{\mathcal{M}}\left(Y(t) \mid \bm{X}(1:t)\right ) = \left( \pi_1 - \pi_0 \right) r_\mathcal{M}(\bm{X}(1:t); \bm{\theta}) + \pi_0. \label{eq:singleConditionalProbabilityExposureModel}
\end{equation*}
In other words, the probability is $\pi_0$ when the patient is at minimal risk ($r_\mathcal{M}(\bm{X}(1:t); \bm{\theta}) = 0$) and $\pi_1$ when the patient is at maximal risk ($r_\mathcal{M}(\bm{X}(1:t); \bm{\theta}) = 1$). This relationship is visualized in Figure~\ref{fig:explanationRelationshipProbabilityRisk}, where the $x$-axis represents the risk level and the $y$-axis signifies the probability of the ADR taking place. As the risk level varies, this probability can assume any value within the interval $[\pi_0, \pi_1]$. Consequently, the conditional probability from equation~\eqref{eq:conditionalDensityFunction} can be expressed as the product
\begin{equation*}
    \p_{\mathcal{M}}\left(\bm{Y} \mid \bm{X} \right)  = \prod_{t = 1}^T \p_{\mathcal{M}}\left(Y(t) \mid \bm{X}(1:t)\right ) = \prod_{t = 1}^T \left[ \left( \pi_1 - \pi_0 \right) r_\mathcal{M}(\bm{X}(1:t); \bm{\theta}) + \pi_0 \right]. 
\end{equation*}
In the next section, we propose eight exposure models that mimic various known relationships between drugs and ADRs. We provide examples of drug-ADR pairs that are known to approximately follow those models. In Section~\ref{sec:parameterEstimation} we discuss how to fit an exposure model $\mathcal{M}$ to a specific data set. It is important to note that the concept of an exposure model is of course not limited to the eight examples presented here; one can define others as well. 

\subsection{Examples of Exposure Models} \label{sec:exampleExposureModels}

Here, we propose eight exposure models by specifying the risk function $r_{\mathcal{M}}$ and the associated parameter space $\Theta_\mathcal{M}$ for each model. 

\subsubsection*{No Assocation ($\mathcal{M}_0$)} 

The no association exposure model reflects the case when there is no association between the drug and ADR in question, i.e., $\p(\bm{X}, \bm{Y}) = \p(\bm{X})\p(\bm{Y})$. In terms of the risk function, we can represent this case as 
\begin{equation*}
    r_{\M_{0}}\left( \bm{X}(1:t) \right) = 0 \qquad \text{for all } \bm{X} \in \mathcal{D}_T. 
    \label{eq:modelNoAssociation}
\end{equation*}
The parameter space $\Theta_{\mathcal{M}_0}$ is the empty set $\emptyset$. In other words, the patient is at `minimal risk' independent of his/her exposure to the drug. See, for example, Figure~\ref{fig:riskModels}a. The drug exposure over time is represented by the $x$-axis, where the exposed period from $t = 5$ to $t = 10$ is denoted by the shaded area. The $y$-axis is the risk level over time given the exposure. As you can see, the risk level is constant and zero throughout. 

\subsubsection*{Current Use $(\mathcal{M}_{\text{current use}})$} 
The current use exposure model represents the case where the patient's risk level is elevated when the patient is exposed and returns to zero the moment he/she ceases to be exposed. Formally, we can express this as 
\begin{equation*}
    r_{\M_{\text{current use}}}\left( \bm{X}(1:t) \right) = X(t) \qquad \text{for all } \bm{X} \in \mathcal{D}_T. 
    \label{eq:modelCurrentUse}
\end{equation*}
The parameter space is $\Theta_{\mathcal{M}_{\text{current use}}} = \emptyset$ as well.
See Figure~\ref{fig:riskModels}b for an example. One can see that the risk level is maximal ($=1$) during exposure and minimal ($=0$) when not exposed. An examples of a drug-ADR pair that, at least approximately, follow this pattern are oral corticosteroids and fractures \citep{vanstaa2000}.

\subsubsection*{Withdrawal Effects $(\mathcal{M}_{\text{withdrawal}})$}

\begin{figure}
    \centering
    \includegraphics[height=.9\textheight]{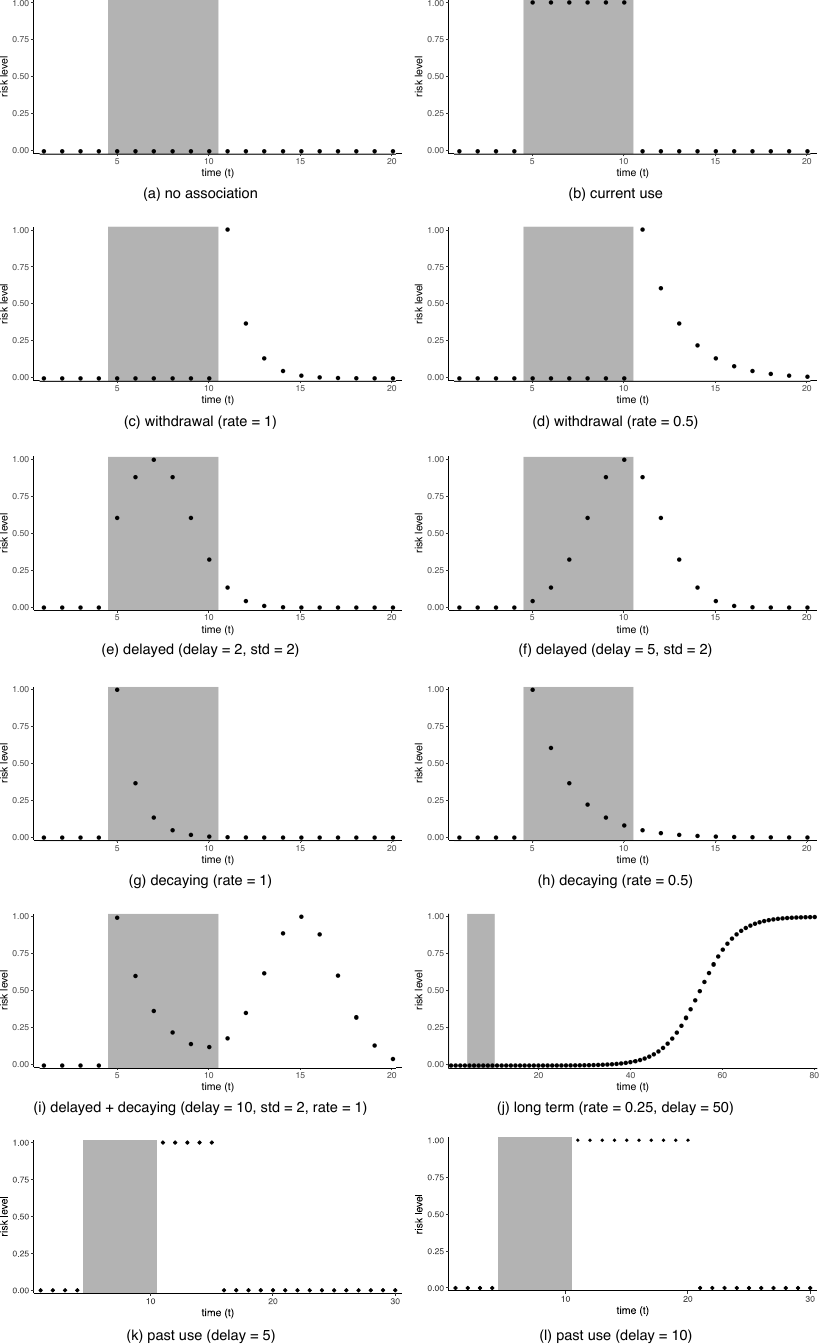}
    \caption{Examples of risk functions introduced in Section~\ref{sec:exampleExposureModels}. These exact risk functions are also used in the simulation study, see Section~\ref{sec:simulationSetUp}.  The horizontal axis represents time ($t$). The gray area is the period in which the patient was exposed (from $t = 5$ to $t = 10$). The $y$-axis shows the risk level, where $1$ represents `maximal' risk to experience the ADR, and $0$ denotes `minimal' risk.}
    \label{fig:riskModels}
\end{figure}

The risk of experiencing withdrawal effects is highest quickly after a patient is no longer exposed and decreases with time. We can model this with the following risk function where, for all $\bm{X} \in \mathcal{D}_T$,
\begin{equation*}
    r_{\M_{\text{withdrawal}}}\left( \bm{X}(1:t); \rho \right) = \begin{cases}
        0 \quad \text{ if never/currently exposed, and} & \\ 
        \exp\left(\rho \cdot f_{\text{last}}(\bm{X}(1:t)) \right) \quad \text{otherwise, }& 
    \end{cases}
    \label{eq:modelWithdrawal}
\end{equation*}
where $\rho \in \Theta_{\mathcal{M}_\text{withdrawal}} = \mathds{R}_+$ is the risk function's only parameter denoting the rate with which the risk level  decreases, and the function $f_\text{last}(\cdot)$ returns
the number of time points since the patient's last exposure, i.e., 
\begin{equation}
    f_\text{last} \left( \bm{X}(1:t) \right) = t - \max\left\{ \tau \in \{1,2,\ldots,t-1\} \text{ such that }X(\tau) = 1 \right\}.
    \label{eq:flast}
\end{equation}
Figure~\ref{fig:riskModels}c and \ref{fig:riskModels}d show two examples where the rate parameter $\rho$ is either $1$ or $\frac{1}{2}$. A well-known example of drugs that elicit such a response are opioids \citep{kosten2019}.

\subsubsection*{Delayed Effects $(\mathcal{M}_{\text{delayed}})$}

The delayed effect model represents drug-ADR pairs where the risk of experiencing the ADR increases gradually during exposure, reaches a `peak' (represented by the parameter $\mu > 0$) and dissipates afterwards. We model this using the probability density function of the normal distribution, normalized so that the risk level is $1$ at $\mu$, i.e., 
\begin{equation}
    r_{\M_{\text{delayed}}}\left( \bm{X}(1:t); \mu, \sigma \right) = \begin{cases}
        0 \quad \text{ if never exposed, and} & \\ 
       \text{exp}\left[-\frac{1}{2}\left(\frac{f_{\text{start}}(\bm{X}(1:t)) - \mu}{\sigma} \right)^2\right]
         \text{ otherwise, }& 
    \end{cases} 
    \label{eq:modelDelayed}
\end{equation}
for all $\bm{X} \in \mathcal{D}_T$. The parameter $\sigma > 0$ regulates how rapidly the risk level increases/decreases and the function  $f_{\text{start}}(\cdot)$ returns the number of time points since the patient was exposed for the first time: 
\begin{equation}
    f_\text{start} \left( \bm{X}(1:t) \right) = t - \min\left\{ \tau \in  \{1,2,\ldots,t\} \text{ such that }X(\tau) = 1 \right\}.
    \label{eq:fstart}
\end{equation}
See Figure~\ref{fig:riskModels}e and Figure~\ref{fig:riskModels}f for an example where $\mu$ is $2$ and $5$, respectively. The parameter $\sigma$ equals $2$ in both cases. Various antiepilectic drugs are known to have a similar temporal relationship to the Stevens-Johnson syndrome \citep{borrelli2018}. Direct oral anticoagulants (DOACs) and gastrointestinal bleeding are also known to follow a similar pattern \citep{garbayo2019}. We consider the latter drug-ADR pair in the case study, see Section~\ref{sec:caseStudy}.

\subsubsection*{Decaying Effects $(\mathcal{M}_{\text{decaying}})$}

In case of a decaying effect, the risk level is maximal when the patient is exposed for the first time and quickly diminishes, even when still exposed. We propose 
\begin{equation*}
    r_{\mathcal{M}_{\text{decaying}}}\left( \bm{X}(1:t); \rho\right) = \begin{cases}
        0 \quad \text{ if never exposed, and} & \\ 
        \exp\left(- \rho \cdot f_{\text{start}}\left(\bm{X}(1:t) \right)\right)
         \text{ otherwise, } &  
    \end{cases} 
    \label{eq:modelDecaying}
\end{equation*}
for all $\bm{X} \in \mathcal{D}_T$, where $\rho > 0$ represents the rate with which the risk level decreases. See Figure~\ref{fig:riskModels}g and \ref{fig:riskModels}h for two examples where the rate $\rho$ is either $1$ or $\frac{1}{2}$. Drug-ADR pairs that are known to follow such a pattern include penicillin and anaphylaxis \citep{neugut2001}, antiepileptic drugs and adverse psychiatric effects \citep{perucca2012, weintraub2007}, and oral contraceptives and venous thrombosis \citep{vanhylckama2009}. We consider penicillin and anaphylaxis in the case study, see Section~\ref{sec:caseStudy}.

\subsubsection*{Delayed and Decaying Effects $(\mathcal{M}_{\text{delayed+decaying}})$}
In some cases, the exposure to a drug shows both a delayed and a decaying effect on the occurrences of the ADR, e.g., oral glucocorticoids and serious infections \citep{dixon2012}. We can reflect this by combining both models as follows:
\begin{equation*}
    r_{\M_{\text{delayed+decaying}}} \left( \bm{X}(1:t); \mu, \sigma, \rho \right) =  
    C^{-1} \big[ r_{\M_{\text{delayed}}}\left( \bm{X}(1:t); \mu, \sigma \right) + r_{\M_{\text{decaying}}}\left( \bm{X}(1:t); \rho \right) \big],
    \label{eq:modelDelayedAndDecaying}
\end{equation*}
where $\mu, \sigma \text{ and } \rho > 0$ and $C$ is a normalizing constant. We must select the value for $C$ such that the maximum value of the model's risk function is $1$. This necessitates solving the optimization problem 
\begin{equation*}
    \begin{split}
        C & = \max_{\bm{X} \in \{0,1 \}^T} \left\{ r_{\M_{\text{delayed}}}\left( \bm{X}(1:t); \mu, \sigma \right) + r_{\M_{\text{decaying}}}\left( \bm{X}(1:t); \rho \right) \right\} \\ & = \max_{\bm{X} \in \{0,1 \}^T} \left\{ \text{exp}\left[-\frac{1}{2}\left(\frac{f_{\text{start}}(\bm{X}) - \mu}{\sigma} \right)^2\right] + \text{exp}\left[- \rho \cdot f_{\text{start}}\left(\bm{X} \right)\right] \right\}. 
    \end{split}
\end{equation*}
To solve this, we would actually need to consider each binary vector $\bm{X} \in \{0,1\}^T$. However, we can simplify the problem using the function $f_\text{start}(\cdot)$. This indicates the number of time points that have elapsed since the patient was first exposed to the drug, see eq.~\eqref{eq:fstart}. 
The values that the function $f_\text{start}(\cdot)$ can take are restricted to ${1,2,\ldots,T - 1}$. Consequently, we can formulate the optimization problem as
\begin{equation*}
    C = \max_{s \in \{1,2,\ldots,T-1\}}  \left\{ \text{exp}\left[-\frac{1}{2}\left(\frac{s - \mu}{\sigma} \right)^2\right] + \exp(- \rho \cdot s ) \right\}, 
\end{equation*}
which can be readily solved numerically. For an illustration of this model, see Figure~\ref{fig:riskModels}i where $\mu = 10$, $\sigma = 2$ and $\rho = 1$. 

\subsubsection*{Long-term Effects $(\mathcal{M}_{\text{long-term}})$}

The ADR can, in some cases, occur long after the patient was exposed for the first time, e.g., antipsychotics and type 2 diabetes \citep{holt2015}. We model these `long-term' cases using a sigmoid function. For all $\bm{X} \in \mathcal{D}_T$, 
\begin{equation*}
    r_{\M_{\text{long-term}}}\left( \bm{X}(1:t); \rho, \kappa \right) = \begin{cases}
        0 \quad \text{ if never exposed, and} & \\ 
        \exp(-\rho (f_{\text{start}}(\bm{X}(1:t)) - \kappa))^{-1}  & \text{otherwise.}
    \end{cases} 
    \label{eq:modelLongTerm}
\end{equation*}
The risk function has two parameters: $(\rho, \kappa ) \in \Theta_{\mathcal{M}_{\text{long-term}}} = \mathds{R}_+^2$. See Figure~\ref{fig:riskModels}j for an example where $\rho = \frac{1}{4}$ and $\kappa = 50$. In contrast to the other figures, the $x$-axis ranges from $t = 1$ to $t = 80$ to illustrate the risk function more clearly. 

\subsubsection*{Past Exposure $(\mathcal{M}_{\text{past}})$}

The risk level of the ADR in question occurring can be elevated from the start of the exposure and remain elevated for a certain period of time, even when the patient is no longer exposed. We model this as
\begin{equation}
    r_{\M_{\text{past}}}\left( \bm{X}(1:t); p\right) = \begin{cases}
        0 & \text{if never exposed, and}  \\ 
        1 & \text{if } f_{\text{last}}(\bm{X}(1:t)) \leq p,
    \end{cases} 
    \label{eq:modelPast}
\end{equation}
for all $\bm{X} \in \mathcal{D}_T$, where $p \in \{1,2,\ldots, T-1\}$ and $f_\text{last}(\cdot)$ is given in eq.~\eqref{eq:flast}. Note that this risk function is the only one proposed here that has a discrete parameter. See Figure~\ref{fig:riskModels}k and \ref{fig:riskModels}l for an example where $p = 5$ and $p = 10$, respectively. Note that in this case, the $x$-axis ranges from $t = 1$ to $t = 30$. 
An example of drug-ADR pair that approximately follows the past exposure model are antiepileptic drugs and delayed allergic hypersensitive reactions \citep{perucca2012,zaccara2007}.

\subsection{Parameter Estimation} \label{sec:parameterEstimation}

In this section, we describe how the parameters of an exposure model $\mathcal{M}$, i.e., $\pi_0$, $\pi_1$ and the parameter vector of the risk function $\bm{\theta}$, can be estimated on the basis of the data. We first consider a general exposure model. Afterwards, we derive the estimators for the no association, current use and past use exposure models as defined in the previous section since, in their case, an analytical solution exists. We denote the data for all patients as $\qunderline{\bm{X}} = \{ \bm{X}^k \}_{k = 1}^N$ and $\qunderline{\bm{Y}} = \{ \bm{Y}^k \}_{k = 1}^N$. The likelihood function for the exposure model $\mathcal{M}$ is given by
\begin{equation*}
    \begin{split}
        \mathcal{L}_{\mathcal{M}}\left( \pi_1, \pi_0, \bm{\theta};  \qunderline{\bm{X}}, \qunderline{\bm{Y}} \right) = \prod_{k = 1}^N \prod_{t = 1}^T & \left[\left( \pi_1 - \pi_0 \right) r_\mathcal{M}\left( \bm{X}^k(1:t); \bm{\theta}\right) + \pi_0 \right]^{Y^k(t)} \times \\ & \qquad \left[ 1 - \left( \pi_1 - \pi_0 \right) r_\mathcal{M}\left( \bm{X}^k(1:t); \bm{\theta}\right) - \pi_0 \right]^{1 - Y^k(t)}.
        \end{split}
\end{equation*}
The corresponding log-likelihood function is then 
\begin{equation}
    \begin{split}
        \ell_{\mathcal{M}}\left( \pi_1, \pi_0, \bm{\theta};  \qunderline{\bm{X}}, \qunderline{\bm{Y}} \right) &  =  \log \mathcal{L}_{\mathcal{M}}\left( \pi_1, \pi_0, \bm{\theta};  \qunderline{\bm{X}}, \qunderline{\bm{Y}} \right)  \\ 
        & = \sum_{k = 1}^N \sum_{t = 1}^T \Big[ Y^k(t) \log\left(\left( \pi_1 - \pi_0 \right) r_\mathcal{M}\left( \bm{X}^k(1:t); \bm{\theta}\right) + \pi_0 \right) + \\ 
        & \qquad \qquad \left(1 - Y^k(t)\right) \log \left( 1 - \left( \pi_1 - \pi_0 \right) r_\mathcal{M}\left( \bm{X}^k(1:t); \bm{\theta}\right) - \pi_0 \right) \Big].
    \end{split}
    \label{eq:loglikelihoodExposureModel}
\end{equation}
Recall that the parameter vector $\bm{\theta} = (\bm{\xi}, \bm{\psi})$ can consist of both continuous ($\bm{\xi}$) and discrete parameters ($\bm{\psi}$). The maximum likelihood estimator (MLE) can, therefore, be written as 
\begin{equation*}
   \begin{split}
        \left(\widehat{\pi}_1, \widehat{\pi}_0, \widehat{\bm{\theta}} \right) &  = \argmax_{\pi_1, \pi_0 \in [0,1], \bm{\xi} \in \Xi_\mathcal{M}, \bm{\psi} \in \Psi_{\mathcal{M}}} \ell_{\mathcal{M}}\left( \pi_1, \pi_0, \left(\bm{\xi}, \bm{\psi} \right);  \qunderline{\bm{X}}, \qunderline{\bm{Y}} \right) \\ & = \argmax_{\bm{\psi} \in \Psi_\mathcal{M}} \left\{ \argmax_{\pi_1, \pi_0 \in [0,1], \bm{\xi} \in \Xi_\mathcal{M}} \ell_{\mathcal{M}}\left( \pi_1, \pi_0, \left(\bm{\xi}, \bm{\psi} \right);  \qunderline{\bm{X}}, \qunderline{\bm{Y}} \right)  \right\}.
   \end{split}
    \label{eq:maximumLikelihoodEstimatorExposureModel}
\end{equation*}
In other words, the original optimization problem can be subdivided into $|\Psi_\mathcal{M}|$ subproblems (where $|\cdot|$ denotes the cardinality of the set), one for each value $\psi \in \Psi_\mathcal{M}$. Analytical solutions for these subproblems might not exist, but they can be solved numerically. We employ Nelder-Mead's algorithm since it allows for discontinuous risk functions as well \citep{nelder1965,avriel2003}.

\subsubsection*{No Association ($\mathcal{M}_0$)}

Recall that the risk function for the null model is $r_{\mathcal{M}_0}(\bm{X}) = 0$ for all $\bm{X} \in \mathcal{D}_T$. Let $Y^+ = \sum_{k = 1}^N \sum_{t = 1}^T Y^k(t)$ be the total number of occurrences of the ADR in the data set. The log-likelihood function reduces to 
\begin{equation*}
   \begin{split}
        \ell_{\mathcal{M}_0}\left( \pi_0; \qunderline{\bm{X}}, \qunderline{\bm{Y}} \right) & = \sum_{k = 1}^N \sum_{t = 1}^T Y^k(t) \log\left(\pi_0\right) + \left(1 - Y^k(t)\right) \log(1 - \pi_0) \\ & = Y^+ \log\left(\pi_0\right) + \left(NT - Y^+ \right) \log\left(1 - \pi_0 \right).
     \end{split}
\end{equation*}
Maximizing this function with respect to $\pi_0$ gives the MLE $\widehat{\pi}_0 = \frac{Y^+}{NT}$.

\subsubsection*{Current use model ($\mathcal{M}_\text{current use}$)}

We can determine the MLE for this model in a similar fashion. Let us first define the following $2 \times 2$ contingency table represented by the random variables $A$, $B$, $C$ and $D$, which are given by
\begin{equation*}
\begin{split}
	A = \sum_{k = 1}^N \sum_{t = 1}^T X^k(t) Y^k(t), \quad B = \sum_{k = 1}^N \sum_{t = 1}^T X^k(t) (1 - Y^k(t)), \quad  \\ 
    C = \sum_{k = 1}^N \sum_{t = 1}^T (1 - X^k(t)) Y^k(t)  \quad \text{and} \quad D = \sum_{k = 1}^N \sum_{t = 1}^T (1 - X^k(t)) (1 - Y^k(t)).
\end{split}
\label{eq:definitionCountsIndividualTimePoints}
\end{equation*}
The count $A$ represent the number of time points the patients were exposed to the drug and experienced the ADR, $B$ is the number of times the patient was exposed, but did not experience the ADR, etc. 
Note that the sum of these counts are the total number of observed time points, i.e., $A + B + C + D = NT$. 
Using these definitions, we can express the exposure model's log-likelihood as 
\begin{equation*}
      \ell_{\mathcal{M}_{\text{current use}}}\left( \pi_1, \pi_0; \qunderline{\bm{X}}, \qunderline{\bm{Y}} \right) = A \log \left( \pi_1 \right) + B \log \left( 1 -  \pi_1 \right) + C \log \left( \pi_0 \right) + D \log \left( 1 - \pi_0 \right).
\end{equation*}
If $A + B > 0$ and $C + D > 0$, the MLE can be calculated as 
$$
    \left( \widehat{\pi}_1, \widehat{\pi}_0 \right) = \left(\frac{A}{A + B}, \frac{C}{C + D} \right).
$$

\subsubsection*{Past Exposure ($\mathcal{M}_\text{past}$)}

In order to derive the MLE for the past use model, we first define the counts $A(p)$, $B(p)$, $C(p)$ and $D(p)$ for $p = 1,2,\ldots,T-1$ as follows: 
\begin{equation*}
    \begin{split}
        A(p) &= \sum_{k=1}^N \sum_{t=1}^T Y^k(t) \mathds{1}\left\{\exists \tau \in \{ \max\{1,t-p\}, \ldots, t\} \text{ such that }X^k(t) = 1 \right\}, \\ 
        B(p) &= \sum_{k=1}^N \sum_{t=1}^T \left(1 -  Y^k(t) \right) \mathds{1}\left\{\exists \tau \in \{ \max\{1,t-p\}, \ldots, t\} \text{ such that }X^k(t) = 1 \right\}, \\
        C(p) &= \sum_{k=1}^N \sum_{t=1}^T Y^k(t)  \mathds{1}\left\{\forall \tau \in \{ \max\{1,t-p\}, \ldots, t\}: X^k(t) = 0 \right\} \text{ and } \\ 
        D(p) &= \sum_{k=1}^N \sum_{t=1}^T \left(1 -  Y^k(t) \right)  \mathds{1}\left\{\forall \tau \in \{ \max\{1,t-p\}, \ldots, t\}: X^k(t) = 0 \right\},
    \end{split}
\end{equation*}
where $A(p)$ denotes the number of occurrences of the ADR when the patient was exposed during the last $p$ time points, $B(p)$ denotes the number of time points when the patient did not experience the ADR, but was exposed during the last $p$ time points etc. We can express the log-likelihood function of this exposure model for a fixed $p$ as 
\begin{equation*}
      \ell_{\mathcal{M}_{\text{past}}}\left( \pi_1, \pi_0, p; \qunderline{\bm{X}}, \qunderline{\bm{Y}} \right) = A(p) \log \left( \pi_1 \right) + B(p) \log \left( 1 -  \pi_1 \right) +  C(p) \log \left( \pi_0 \right) + D(p) \log \left( 1 - \pi_0 \right)
\end{equation*}
which gives the following MLE of $(\pi_1, \pi_0)$, if $A(p) + B(p) > 0$ and $C(p) + D(p) > 0$: 
$$
    \left( \widehat{\pi}_1, \widehat{\pi}_0 \right) = \left(\frac{A(p)}{A(p) + B(p)}, \frac{C(p)}{C(p) + D(p)} \right).
$$
Using the result in eq.~\eqref{eq:loglikelihoodExposureModel}, the estimator for the past use model can be expressed as 
\newcommand{\CP}[3]{#1(p) \log\left(\frac{#1(p)}{#2(p) + #3(p)} \right)}
\begin{equation*}
    \begin{split}
    \left( \widehat{\pi}_1, \widehat{\pi}_0, \widehat{p} \right) & = \argmax_{p \in \{1,2\ldots,T-1\}} \left\{ \argmax_{\pi_1, \pi_0 \in [0,1]} \ell_{\mathcal{M}_\text{past}} \left( \pi_1, \pi_0, p;  \qunderline{\bm{X}}, \qunderline{\bm{Y}} \right)  \right\} \\ 
    & = \argmax_{p \in \{1,2\ldots,T-1\}} \Bigg\{ \CP{A}{A}{B} + \CP{B}{A}{B} +  \\ 
    & \qquad \qquad \qquad \quad \CP{C}{C}{D} + \CP{D}{C}{D} \Bigg\}. 
    \end{split}
\end{equation*}

\subsection{Model Selection} \label{sec:modelSelection}

There is a variety of model selection approaches available in the literature that can aid in selecting the `best' exposure model after fitting it to the data \citep{ding2018}. Here we opt for the Bayesian Information Criterion (BIC) due to its connection to the posterior probabilities of the models  \citep{schwarz1978}. The BIC for an exposure model $\mathcal{M}$ is given by 
\begin{equation*}
    \text{BIC} = (q + 2) \log(N) - 2\ell_\mathcal{M} \left( \pi_1^\ast, \pi_0^\ast, \bm{\theta}^\ast; \qunderline{\bm{x}}, \qunderline{\bm{y}} \right),
    \label{eq:definitionBIC}
\end{equation*}
where $q$ is the number of parameters of the risk function, i.e., the dimensionality of the parameter space $\Theta_\mathcal{M}$, $\qunderline{\bm{x}} = \{ \bm{x}^k \}_{k = 1}^N$ and $\qunderline{\bm{y}} = \{ \bm{y}^k \}_{k = 1}^N$ are the observed data, and $\pi_1^\ast$, $\pi_0^\ast$ and $\bm{\theta}^\ast$ are the values for which the log-likelihood is maximal. Note that with the BIC, one tries to strike a balance between the fit (log-likelihood) and the model's complexity expressed by the total number of parameters $(q + 2)$. 

Suppose we consider the exposure models $\mathcal{M}_1, \mathcal{M}_2, \ldots, \mathcal{M}_V$, where $V = 8$ in our case, and let $\text{BIC}_v$ be the BIC-score for the $v$-th model. The model with the lowest BIC-score is preferred. Schwarz\citep{schwarz1978} shows that the posterior probability of the $v$-th model can be approximated by 
\begin{equation*}
    \p\left(\mathcal{M}_v \mid \qunderline{\bm{x}}, \qunderline{\bm{y}} \right) \approx \frac{\exp\left( - \frac{1}{2} \text{BIC}_v \right)}{\sum_{w = 1}^V \exp\left( - \frac{1}{2} \text{BIC}_w \right)}.
\end{equation*}
This result is especially useful in our case for two reasons. First, one might be interested in determining whether there is an association between the drug and ADR in question and not in identifying which exposure model fits the data `best' in itself. To this end, one can use the posterior probability of the no association exposure model. For example, if the posterior probability $\p(\mathcal{M}_0 \mid \qunderline{\bm{x}}, \qunderline{\bm{y}}) \geq \frac{1}{2}$, there is no association and the drug-ADR pair is not reported as a signal. 

The second reason why the use of the posterior probabilities is convenient, is its ability to facilitate a comparison of various drug-ADR pairs and establish a ranking from `interesting' to less `interesting'. Let $\qunderline{\bm{x}_i}$ and $\qunderline{\bm{y}_j}$ be the observed data for the \ith drug and the \jth ADR, respectively, and let $\p(\mathcal{M}_0^{ij} \mid \qunderline{\bm{x}_i}, \qunderline{\bm{y}_j})$ be the posterior probability of the null model for the drug-ADR pair $(i,j)$. We can create a ranking of pairs based on these scores, where the lower the posterior probability is, the stronger the signal for the pair is deemed to be. An advantage of this approach is that one can employ Bayesian false discovery rate control procedures, see, for example, the work by Storey\citep{storey2003}, to determine which signals to present to the committee of medical experts.

\section{Simulating Electronic Healthcare Data} \label{sec:simulatingEHCData} 

In this section, we describe the simulator for EHC data used for the simulation study. We start with how we model drug exposures over time, after which we show how to generate ADR occurrences given the drug exposure history and an exposure model. We finish with the pseudo-algorithm. 

\subsection*{Drug Exposures} \label{sec:simulatingDrugExposures}
We model the exposure of a patient to the drug of interest over time as a Markov chain. Let $\bm{X} = \left(X(1), X(2), \ldots, X(T)\right)$ be the binary time series. We then specify the Markov chain by 
\begin{equation}
    \p(X(1) = 1) = \nu_0 \quad \text{and} \quad \p(X(t) = 1 \mid X(t - 1) = x) = \begin{cases}
        \nu_0 & \text{if }x = 0\\ 
        \nu_1 & \text{if }x = 1
    \end{cases} 
    \label{eq:markovChainDefinition}
\end{equation}
for $t = 2,3, \ldots, T$. Rather than thinking in terms of the probabilities $\nu_0$ and $\nu_1$, we find it more intuitive to consider 1) the probability of the patient to be exposed to the drug at least once, and 2) the average duration of the exposure once exposed. Let $E$ be a binary random variable representing whether the patient was exposed at least once ($E = 1$) or not ($E = 0$), and let $D \in \mathds{N}$ be a random variable denoting the duration of the exposure once exposed. We find that the probability of being exposed can be expressed as 
\begin{equation*}
        \mu_{E}   = \p(E = 1) = 1 - \p(E = 0)  = 1 - \p(X(1) = 0, X(2) = 0, \ldots, X(T) = 0) = 1 - \left(1 - \nu_0\right)^T.  
    \label{eq:definitionProbabilityPatientBeingExposed}
\end{equation*}
So, rather than choosing $\nu_0$ directly, we select the probability of a patient to be exposed ($\mu_{E}$) and set $\nu_0 = 1 - \left(1 - \mu_{E}\right)^{\frac{1}{T}}$. 

Once a patient is exposed, the probability that he/she is exposed at the following time point is $\nu_1$. The duration $D$ of the exposure, therefore, follows a geometric distribution with probability density function 
\begin{equation*}
    \p(D = d) = \left( \nu_1 \right)^{d} \left(1 - \nu_1 \right)
\end{equation*}
with mean 
\begin{equation*}
    \delta = \mathds{E}(D) = (1 - \nu_1)^{-1}.
\label{eq:definitionAverageDurationExposure}
\end{equation*}
Once one chooses the average duration of the exposure, $\delta$, one can set $\nu_1 = (\delta - 1) / \delta$. In our simulation set-up, we choose an average duration of $\delta = 5$ time points, i.e., $\nu_1 = .8$. 

\subsection*{Adverse Drug Reactions} \label{sec:simulatingADRs}

Let $\bm{X} = \bm{x} = \left(x(1), x(2), \ldots, x(T) \right)$ be a simulated drug history. Given an exposure model $\mathcal{M}$ with risk function $r_\mathcal{M}(\cdot; \bm{\theta})$ and probabilities $\pi_0$ and $\pi_1$, the random binary variable $Y(t)$ follows, conditional on the drug history, a Bernoulli distribution, i.e., 
\begin{equation}
    Y(t) \mid \bm{X}(1:t) = \bm{x}(1:t) \sim \text{Bernoulli}\left((\pi_1 - \pi_0) r_{\mathcal{M}}(\bm{x}(1:t); \bm{\theta}) + \pi_0 \right).
    \label{eq:bernoulliDistributionForADR}
\end{equation}

\subsection*{Pseudo-Algorithm} \label{sec:pseudoAlgorithm}
Combining these steps, we propose the following procedure for generating EHC data for a single drug-ADR pair: 
\begin{enumerate}
    \item Select the number of patients ($N$), an exposure model $\mathcal{M} = \left<\pi_1, \pi_0, T, r_{\mathcal{M}}(\cdot; \bm{\theta}), \Theta \right>$, the probability of a patient to be exposed $(\mu_{E})$, and the average duration of the exposure once exposed ($\delta$);
    \item Determine the probabilities $\nu_0 = 1 - (1 - \mu_{E})^{\frac{1}{T}}$ and $\nu_1 = (\delta - 1)/\delta$ governing the Markov chain in eq.~\eqref{eq:markovChainDefinition};
    \item{ For all patients $k = 1,2,\ldots,N$ perform the following two steps: 
        \begin{enumerate}
            \item Sample a drug history for patient $k$ according to eq.~\eqref{eq:markovChainDefinition}. We denote the resulting drug history by $\bm{x}^k$, and 
            \item Generate the ADR history $\bm{y}^k = (y^k(1), y^k(2), \ldots, y^k(T) )$ for patient $k$ by sampling from the Bernoulli distribution in eq.~\eqref{eq:bernoulliDistributionForADR} given the drug history $\bm{x}^k$ from the previous step.
        \end{enumerate}
    }
\end{enumerate}
An implementation of this algorithm is publicly available as an \texttt{R}~package at {\url{https://github.com/bips-hb/expard}.

\section{Simulation Set-Up} \label{sec:simulationSetUp}

In our simulation, we simulate one drug-ADR pair at the time following the procedure as described in the previous section. The drug-ADR pair can follow one out of twelve exposure models. We list them here. For a visual representation, see Figure~\ref{fig:riskModels}; 
\begin{enumerate}
    \item no association; 
    \item current use;
    \item a withdrawal effect with a rate of $\rho = 1$; 
    \item a withdrawal effect with a rate of $\rho = \frac{1}{2}$; 
    \item a delayed effect with its `peak' at time point $\mu = 2$, and $\sigma = 2$; 
    \item a delayed effect with its `peak' at time point $\mu = 5$, and $\sigma = 2$; 
    \item a decaying effect with a rate of $\rho = 1$; 
    \item a decaying effect with a rate of $\rho = \frac{1}{2}$; 
    \item a combination of a delayed and decaying effect where $(\mu, \sigma, \rho) = (10,2,1)$; 
    \item a long-term effect with rate $\rho = \frac{1}{4}$ and $\kappa = 50$; 
    \item a past use model with $p = 5$, and 
    \item a past use model with $p = 10$. 
\end{enumerate}
The number of patients and time points are fixed at $N = 1,000$ and $T = 100$, respectively. The probability for a patient to be exposed to the drug ($\mu_{E}$) is varied from $.01$, $.1$ to $.5$. The average duration of an exposure once exposed ($\delta$) is 5 time points. The probability $\pi_1$ is $.01$, $.1$, $.2$ or $.3$; the probability $\pi_0$ is either $10^{-4}$ or $10^{-3}$. See Table~\ref{tab:simulationParametersEHC} for an overview of all the parameter settings. We, thus, consider a total of 288 parameter settings. We repeat the simulation for each setting 20 times. 

\begin{table}[h!]
    \centering
    \caption{Parameter settings used in the EHC simulation study}
    \begin{tabular}{l c c l} 
        \toprule
         \textbf{Description} &  \textbf{Notation} & \textbf{Values} \\ 
         \midrule
         Number of patients &  $N$ & 1000 \\ 
         Number of time points &  $T$ & 100 \\ 
         \midrule
         Probability being exposed &  $\mu_{E}$ & $.01, .1$ or $.5$ \\
         Average duration exposure & $\delta$ & 5 \\ \midrule
         Probability ADR minimal risk & $\pi_0$ & $10^{-4}$ or $10^{-3}$  \\
         Probability ADR maximal risk & $\pi_1$ & $.01, .1, .2$ or $.3$  \\
         \bottomrule
    \end{tabular}
    \label{tab:simulationParametersEHC}
\end{table}

\section{Performance Assessment} \label{sec:performanceAssessment} 

We assess the method's effectiveness by considering two aspects: 1) its ability to accurately identify the correct exposure model (model selection), and 2) its capacity to ascertain the presence or absence of an association between the drug and the ADR (signal detection). Although these two tasks are related, it is important to note that it might be of interest to understand the nature of the relationship between the drug and ADR or to simply confirm the existence of a relationship. In the following two sections, we detail how we measure the method's performance with regard to these aspects. 

\subsection{Model Selection} \label{sec:performanceAssessmentModelSelectionConfusionMatrices}

We consider two approaches for model selection. We either select 1) the model with the lowest BIC (or, conversely, the highest posterior probability), see Section~\ref{sec:modelSelection}, or 2) the model with the highest likelihood. The former approach considers the complexity of the models in terms of the number of parameters, while the latter solely focuses on the model's fit.

For every combination of $\mu_E$, $\pi_0$, and $\pi_1$ (see Section~\ref{sec:simulationSetUp}), we evaluate the performance by examining the \emph{confusion matrix}. Examples of such a confusion matrix can be found in Figures~\ref{fig:confusionMatrixPerfectScore},  \ref{fig:resultsSimulationExposure001} and \ref{fig:resultsSimulationExposure05}. 
The horizontal axis represents the twelve true exposure models (see Section~\ref{sec:simulationSetUp} for an overview), while the vertical axis indicates the model selected based on the BIC. The values within each cell indicate the number of times the respective selected model was chosen when the data was simulated based on the corresponding true model. For readability, cells with the value $0$ are left empty. The difference between the number of true models (twelve) and selectable models (eight) is due to the fact that true models serve as the foundation for the simulation with predetermined parameters, whereas the parameters of the chosen model are estimated based on the data (see Section~\ref{sec:parameterEstimation}).
Note that each column in the confusion matrix always adds up to 20, corresponding to the number of repetitions. 
To clarify, Figure~\ref{fig:confusionMatrixPerfectScore} presents the ideal confusion matrix, where each true model is correctly identified. For example, withdrawal models with rates 1 and $\frac{1}{2}$ are both identified as the withdrawal model.  

\begin{figure}[h]
    \centering
    \includegraphics[width=.7\textwidth]{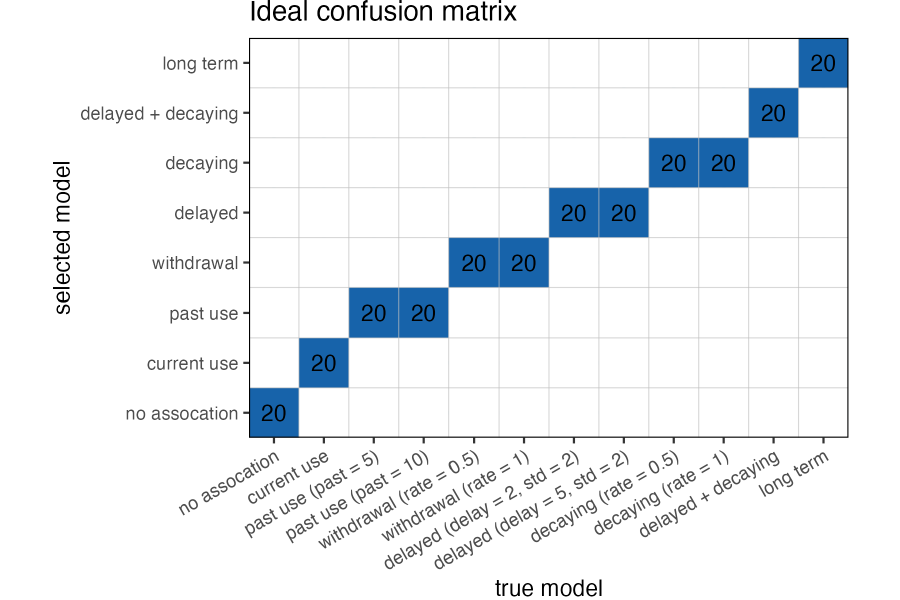}
    \caption{The confusion matrix in the ideal scenario where the method classifies each case perfectly. The twelve true models utilized in the simulation study (see Section~\ref{sec:simulationSetUp}) are represented along the horizontal axis. On the vertical axis, the exposure model that was selected is shown. Note that, for instance, choosing `withdrawal' is correct for both true withdrawal models where the rate is either 1 or $\frac{1}{2}$. The number 20 denotes the total number of repetitions.}
    \label{fig:confusionMatrixPerfectScore}
\end{figure}

\subsection{Signal Detection}

Our focus here lies in evaluating the the method's performance in detecting the presence or absence of an association between the drug and ADR of interest. Recall that we simulate twelve exposure models for every combination of $\mu_E$, $\pi_0$, and $\pi_1$, as outlined in Section~\ref{sec:simulationSetUp}. We repeat this process $20$ times, resulting in a total of $12 \cdot 20 = 240$ runs. 
A drug-ADR pair is not associated when the true model is the no association model $\mathcal{M}_0$, and associated if the true model is one of the remaining eleven. To frame this as a binary classification problem, we define two sets: let $\bm{I}^\ast = \{I_1^\ast, I_2^\ast, \ldots, I_{240}^\ast\}$ represent the truth, where $I_l^\ast = 0$ if, for the $l$-th run, the true model is the null model, and $1$ otherwise. Note that only $20$ values in $\bm{I}^\ast$ are 0. Let $\bm{I} = \{I_1, I_2, \ldots, I_{240}\}$ represent the method's decisions, where $I_l$ is 0 if the method indicates no association, and 1 otherwise. We can then define the number of true positives (TP), true negatives (TN), false positives (FP), and false negatives (FN) as
 \begin{equation*}
\begin{split}
	\text{TP} = \sum_{l = 1}^{240} I_l^\ast I_l, \quad \text{TN} = \sum_{l = 1}^{240} \left(1 - I_l^\ast\right) \left(1 - I_l\right), \quad \\ 
    \text{FP} = \sum_{l = 1}^{240} \left( 1 - I_l^\ast \right) I_l \quad \text{ and } \quad \text{FN} = \sum_{l = 1}^{240} I_l^\ast \left( 1 - I_l \right).
\end{split}
\label{eq:definitionTPTNFPFN}
\end{equation*}
In situations involving unbalanced data, as is the case here, it is recommended to use precision and recall as performance measures \citep{saito2015}. These metrics are defined as
\begin{equation}
    \text{Precision} = \frac{\text{TP}}{\text{TP} + \text{FP}} \quad \text{and} \quad \text{Recall} = \frac{\text{TP}}{\text{TP} + \text{FN}}. 
    \label{eq:definitionPrecisionRecall}
\end{equation}
The $F_1$ score is their harmonic mean, i.e., 
\begin{equation}
    F_1 = 2 \cdot \frac{\text{Precision} \cdot \text{Recall}}{\text{Precision} + \text{Recall}}. 
    \label{eq:definitionF1Score}
\end{equation}
To highlight the impact of using either the posterior probability (linked to the BIC) or the likelihood directly, we decide on the presence of an association between the drug and ADR based on whether: 1) the posterior probability of the null model exceeds $.5$, or 2) the model with the highest likelihood is not the null model.

\section{Case Study} \label{sec:caseStudy}


To demonstrate our exposure model framework, we implement the suggested approach using data from the German Pharmacoepidemiological Research Database (GePaRD; \citep{haug2021}). The data is from two statutory health insurance (SHI) providers in Germany: hkk Krankenkasse and AOK Bremen/Bremerhaven. Our analysis focuses solely on in-patient data and individuals who were insured 1) during the years 2004 until 2017, and 2) for a consecutive period of time, allowing for a maximum gap of 14 days. The total number of patients considered exceeds 1.2 million. The temporal resolution is set to quarter years. 



We consider four drug-ADR pairs for which the temporal relationship is known in the literature:
\begin{enumerate}
    \item Penicillin (ATC: J01C) and anaphylaxis (ICD: T88.6): In the case of a penicillin allergy, the reaction typically occurs almost immediately, with the associated risk diminishing rapidly over time. This aligns closely with the decaying exposure model  \citep{neugut2001}. However, it is important to note that detecting this pattern requires high time resolution (days rather than quarters). Consequently, the current use model is anticipated to offer a more fitting description of the relationship;
    \item Direct oral anticoagulants (DOACs; ATC: B01AF) and gastrointestinal (GI) bleeding\footnote{The ICD-codes related to gastrointestinal (GI) bleeding are: I98.3, K22.6, K22.8, K22.80, K22.81, K22.88, K25.0, K25.2, K25.4, K25.6, K26.0, K26.2, K26.4, K26.6, K27.0, K27.2, K27.4, K27.6, K28.0, K28.2,K 28.4, K28.6, K29.0, K31.8, K55.2, K55.3, K55.8, K57.0, K57.1, K57.2, K57.3, K57.4, K57.5, K57.8, K57.9, K62.5, K66.1, K92.0, K92.1, K92.2.}: The probability of bleeding increases with prolonged exposure, reaches a `peak' and decreases afterwards. This pattern aligns most closely with the delayed exposure model \citep{garbayo2019};
    \item Antipsychotics (ATC: N05A) and type 2 diabetes (ICD: E11): Prolonged use of antipsychotics has been associated with weight gain and the onset of type 2 diabetes. The long-term exposure model seems to be most appropriate \citep{holt2015}, and
    \item Antibiotics (ATC: J01) and GI bleeding: This combination is our negative control, as there is no evidence that the use of antibiotics elevates the risk of GI bleeding. We incorporate this negative control to evaluate the effectiveness of our method in identifying the null model.
\end{enumerate}

\section{Results} \label{sec:results}

We present the results from both the simulation and the case study. 

\subsection{Simulation Study}
\label{sec:resultsSimulationStudy}

Initially, we investigate the method's capability to select the true exposure model. Subsequently, we evaluate its effectiveness in accurately determining the existence of an association between the drug and the ADR.

\subsubsection{Model Selection}
In this section, we show the simulation results for a select number of parameter settings. You can interactively explore the results for the other parameter settings at \url{https://exposuremodels.bips.eu}. The trends shown here in this section are applicable to the broader range of parameter settings as well.

Figures~\ref{fig:resultsSimulationExposure001} and \ref{fig:resultsSimulationExposure05} display a set of confusion matrices (see Section~\ref{sec:performanceAssessmentModelSelectionConfusionMatrices}), where the probability of being exposed to the drug at least once ($\mu_E$) is, respectively, relatively low, i.e., around $1\%$ of the patients, and high, i.e., around $50\%$. Both figures are organized as follows: rows represent different values of $\pi_0$, signifying the probability of experiencing the ADR when the patient is at minimal risk (a risk level of $0$). The top and bottom rows correspond to $\pi_0 = 10^{-4}$ and $\pi_0 = 10^{-3}$, respectively. Columns represent different values of $\pi_1$, denoting the probability of the ADR occurring when the patient is at maximal risk (a risk level of $1$). The left and right columns correspond to $\pi_1 = .01$ and $\pi_1 = .3$, respectively.

Figure~\ref{fig:resultsSimulationExposure05} illustrates the simulation results when the probability of exposure is high ($\mu_E = .5$). The figure follows the same structure as Figure~\ref{fig:resultsSimulationExposure001}, with the rows presenting the results for $\pi_0$ values of $10^{-4}$ and $10^{-3}$, respectively, and with the columns containing the results for $\pi_1$ values of $.01$ and $.3$, in that order. 

\begin{sidewaysfigure}
    \centering
    \includegraphics[width=.8\textwidth]{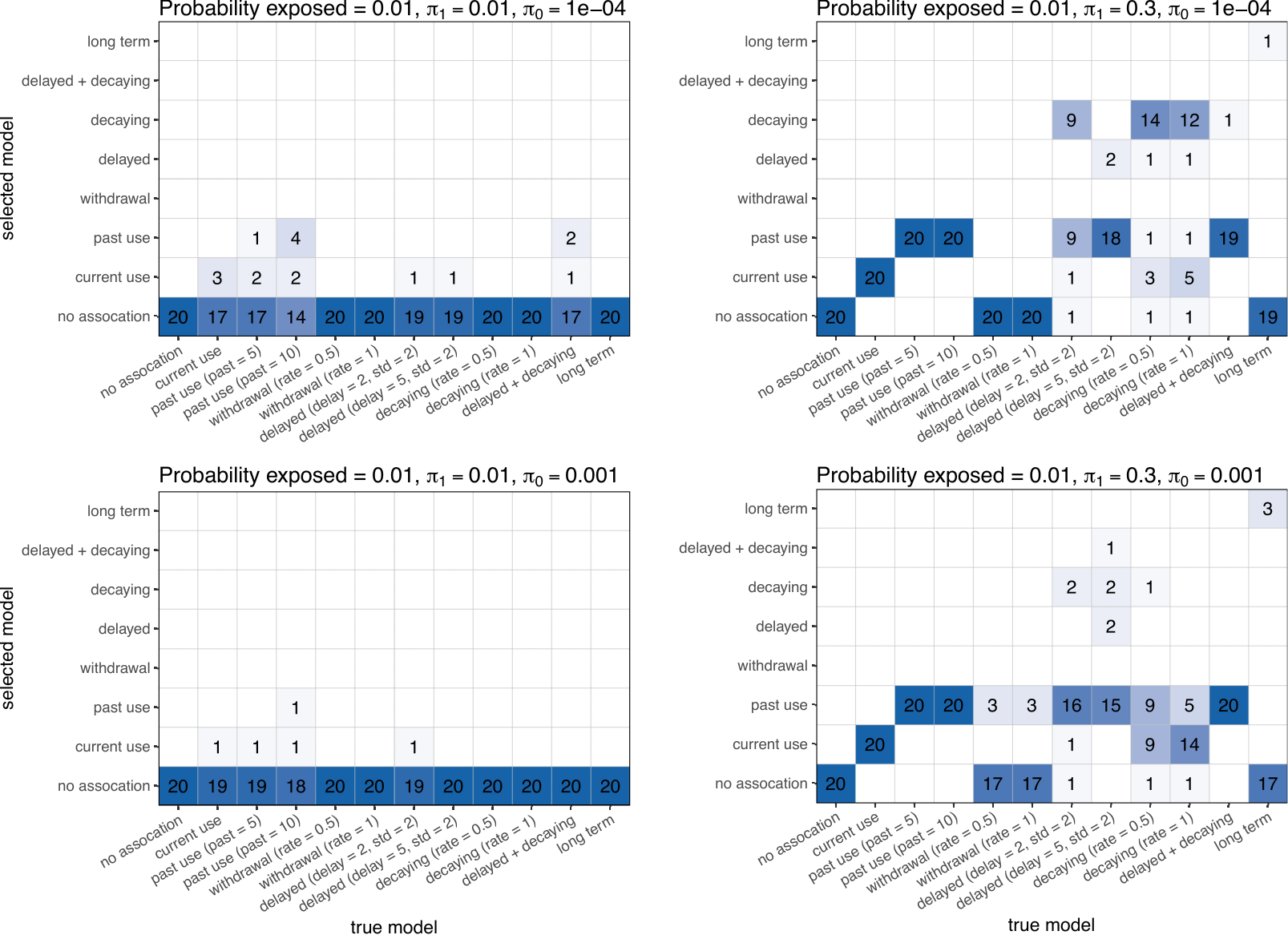}
    \caption{Several confusion matrices showing the simulation results when the number of exposed patients is low, approximately 1\% of the patients. The $x$-axes represent the twelve true models, while the $y$-axes indicate the models selected based on the BIC. The top and bottom rows illustrate the results when $\pi_0$ is $10^{-4}$ and $10^{-3}$, respectively. The left and right columns present the outcomes for $\pi_1$ values of $.01$ and $.3$, respectively.}
    \label{fig:resultsSimulationExposure001}
\end{sidewaysfigure}

\begin{sidewaysfigure}
    \centering
    \includegraphics[width=.8\textwidth]{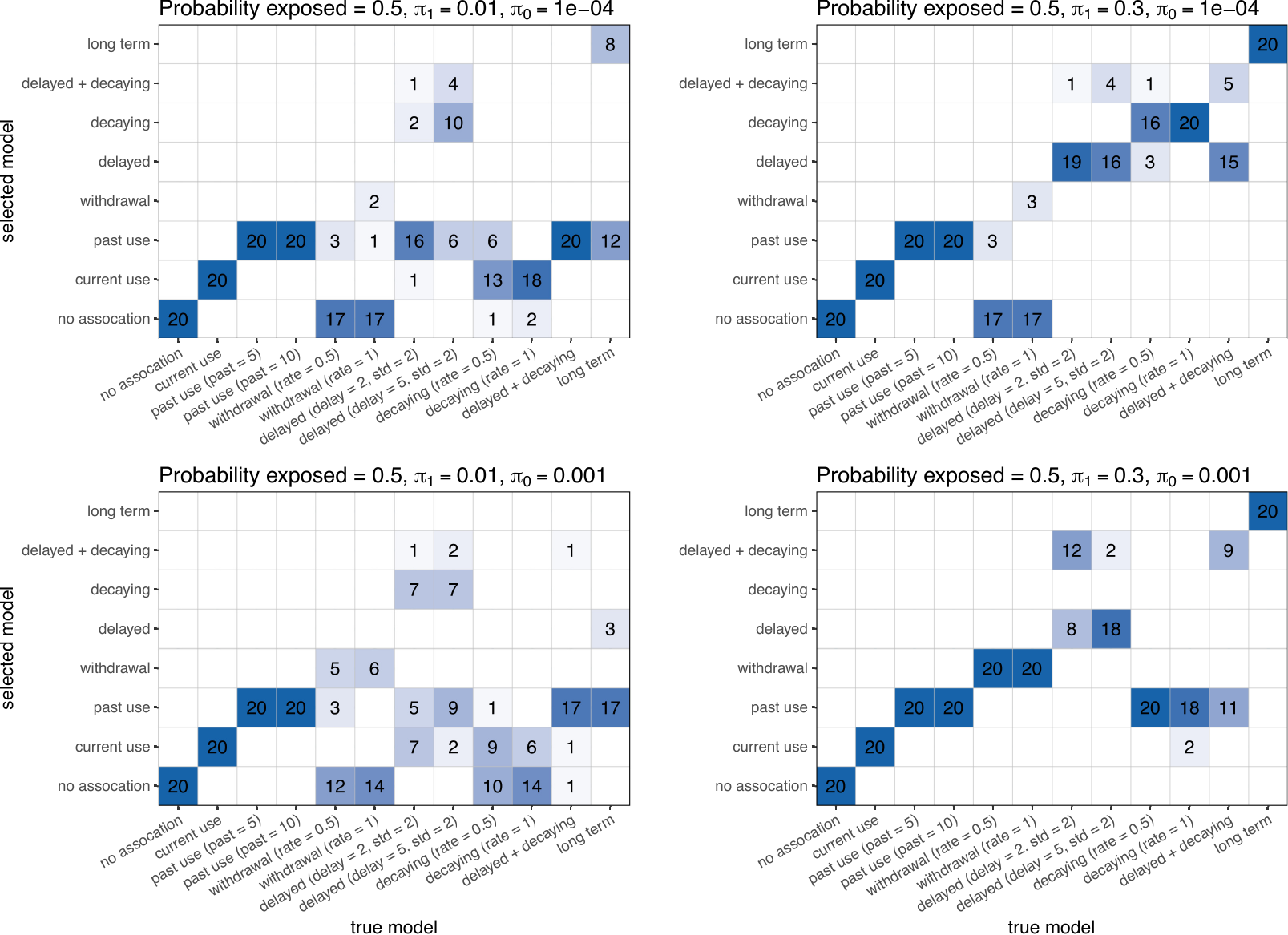}
    \caption{Several confusion matrices showing the simulation results when the number of exposed patients is high, approximately 50\% of the patients. The $x$- and $y$-axes denote the true and selected models, respectively. The top and bottom rows illustrate the results with $\pi_0$ set at $10^{-4}$ and $10^{-3}$, while the left and right columns present the outcomes for $\pi_1$ values of $.01$ and $.3$, in that order.}
    \label{fig:resultsSimulationExposure05}
\end{sidewaysfigure}

The lowest performance observed shows the confusion matrix in the lower left corner of the figure. Two factors contribute to the poor performance: 1) the probability of experiencing the ADR at maximal risk, denoted as $\pi_1$, is relatively low ($\pi_1 = .01$), and 2) the difference between the probabilities of experiencing the ADR at maximal and minimal risk, represented as $|\pi_1 - \pi_0|$, is smaller compared to the top row of the figure. A reduced disparity between $\pi_1$ and $\pi_0$ makes it more challenging to detect an association between the drug and the specific ADR. While the performance in the top left corner is slightly better, reliably identifying a signal under these conditions remains impossible. The suboptimal results in the left column of the figure are not surprising; only a limited number of patients are exposed, and even
when individuals are at risk, occurrences of the ADR are infrequent.

The outcomes in the right column of Figure~\ref{fig:resultsSimulationExposure001} show significant improvement, mainly because the difference $|\pi_1 - \pi_0|$ is considerably larger than in the left column. The probability of the ADR occurring is significantly higher when the patient is at risk, making it easier to accurately identify the exposure model.

The exposure models delayed or delayed + decaying, see Section~\ref{sec:exampleExposureModels}, are frequently misclassified as the past use model. This misclassification arises due to the fact that these models share important characteristics; namely, the delayed, delayed + decaying, and past use models all indicate that the patient is at an increased risk of experiencing the ADR during or after being exposed to the drug, see Figure~\ref{fig:riskModels}. Given our use of the BIC for model selection, the past use model is favored due to its fewer parameters -- three as opposed to four and five parameters for the delayed and delayed + decaying models, see Section~\ref{sec:exampleExposureModels}. The available data do not provide sufficient support for differentiating between these three models. However, this scenario changes in Figure~\ref{fig:resultsSimulationExposure05} where the number of exposed patients is significantly higher.

For most exposure models, detection improves as the difference $|\pi_1 - \pi_0|$ increases. However, this trend does not apply to the withdrawal model, which, surprisingly, becomes easier to detect when the difference decreases slightly. Although this effect is less pronounced in this figure, it becomes more evident in Figure~\ref{fig:resultsSimulationExposure05}. 

Detecting long-term models proves to be very challenging under the settings considered. 

Let us consider the results in Figure~\ref{fig:resultsSimulationExposure05} when the probability of exposure is high ($\mu_E = .5$). 
The performance significantly improves compared to the scenario depicted in Figure~\ref{fig:resultsSimulationExposure001}. This improvement is expected since approximately half of the patients are now exposed, a substantial increase from the previous 1\%. Similarly, we observe that the performance is influenced by the difference between the probability of the ADR occurring when the patient is at maximal or minimal risk, i.e., $|\pi_1 - \pi_0|$. Specifically, the top row, where this difference is larger, exhibits better performance than the bottom row. 

Furthermore, the delayed and decaying models and their combination are still frequently misclassified as the past use model for the same reason mentioned earlier. The optimal performance occurs in the upper right corner, indicating that under these conditions, the data supports choosing these models over the past use model. The withdrawal model is easier to detect when the difference $|\pi_1 - \pi_0|$ is smaller, contrary to the other models. However, the long-term model remains challenging to detect when $\pi_1$ is $.01$ (left column) but becomes detectable when $\pi_1$ is $.3$ (right column).

\subsubsection{Signal Detection}

The results are presented in Table~\ref{tab:performanceResultsBICLogLikelihood}, where the first three columns display the parameter settings, the next three columns showcase the results when the posterior probability is utilized, and the last three columns illustrate the outcomes when the decision is based on the likelihood of the model. 

Similar patterns emerge here as those observed in the previous section on model selection. Performance tends to be poor when the number of patients exposed at least once is low, improving as this number increases. As the probability of experiencing the ADR at maximal risk ($\pi_1$) increases, performance also improves, aligning with expectations. 
Comparing cases where $\mu_E$ and $\pi_1$ remain constant show better performance when $\pi_0 = 10^{-4}$ is low, since the difference between $\pi_1$ and $\pi_0$ is larger.

Utilizing the posterior probability results in overly conservative decisions, as evidenced by a precision of 1 for all parameter settings, indicating a reluctance to produce a signal. The use of maximum likelihood improves the situation, although precision may decrease in some instances. Notably, when there are more patients exposed or the frequency of the ADR increases, the performance of the method based on the posterior probability surpasses that of the maximum likelihood-based approach.

\begin{table}[h!]
\centering
\caption{The performance when either the posterior probability or the maximum likelihood is used for determining whether there is an association between the drug and ADR in question. The parameters used in the simulation are the probability to be exposed ($\mu_E$) and the probabilities for the ADR to occur when the patient is at minimal ($\pi_0$) or maximal risk ($\pi_1$). See for the definitions of the precision, recall and $F_1$ score eq.~\eqref{eq:definitionPrecisionRecall} and \eqref{eq:definitionF1Score}.}
\small
\begin{tabular}{c c c | c c c | c c c}
 \toprule
 \multicolumn{3}{c}{\textbf{Parameter settings}} & \multicolumn{3}{c}{\textbf{Posterior probability}} & \multicolumn{3}{c}{\textbf{Max. likelihood}} \\ 
 $\mu_E$ & $\pi_0$ & $\pi_1$ & \textbf{Precision} & \textbf{Recall} & $F_1$ & \textbf{Precision} & \textbf{Recall} & $F_1$ \\
  \midrule
  .01 & $10^{-4}$ & .01 & 1 & .08 & .14 & 1 & .22 & .36 \\ 
  .01 & $10^{-4}$ & .10 & 1 & .50 & .67 & 1 & .58 & .74 \\ 
  .01 & $10^{-4}$ & .20 & 1 & .70 & .82 & 1 & .71 & .83 \\ 
  .01 & $10^{-4}$ & .30 & 1 & .72 & .84 & 1 & .73 & .85\\ 
  \midrule
  .01 & $10^{-3}$ & .01 & 1 & .02 & .04 & .95 & .27 & .42 \\  
  .01 & $10^{-3}$ & .10 & 1 & .45 & .62 & .98 & .62 & .76 \\  
  .01 & $10^{-3}$ & .20 & 1 & .69 & .81 & .98 & .75 & .85 \\  
  .01 & $10^{-3}$ & .30 & 1 & .75 & .86 & .98 & .77 & .86 \\  
  \midrule
  .10 & $10^{-4}$ & .01 & 1 & .53 & .69 & 1 & .64 & .78 \\ 
  .10 & $10^{-4}$ & .10 & 1 & .75 & .86 & 1 & .75 & .86 \\ 
  .10 & $10^{-4}$ & .20 & 1 & .77 & .87 & 1 & .77 & .87 \\ 
  .10 & $10^{-4}$ & .30 & 1 & .78 & .88 & 1 & .78 & .88 \\ 
  \midrule
  .10 & $10^{-3}$ & .01 & 1 & .27 & .43 & .95 & .77 & .85 \\ 
  .10 & $10^{-3}$ & .10 & 1 & .87 & .93 & .95 & .87 & .91 \\ 
  .10 & $10^{-3}$ & .20 & 1 & .87 & .93 & .96 & .87 & .91 \\ 
  .10 & $10^{-3}$ & .30 & 1 & .88 & .93 & .96 & .88 & .91 \\ 
  \midrule
  .50 & $10^{-4}$ & .01 & 1 & .83 & .91 & .98 & .85 & .91 \\ 
  .50 & $10^{-4}$ & .10 & 1 & .85 & .92 & .98 & .85 & .91 \\ 
  .50 & $10^{-4}$ & .20 & 1 & .85 & .92 & .98 & .85 & .91 \\ 
  .50 & $10^{-4}$ & .30 & 1 & .85 & .92 & .98 & .85 & .91 \\ 
  \midrule
  .50 & $10^{-3}$ & .01 & 1 & .77 & .87 & .92 & 1 & .96 \\ 
  .50 & $10^{-3}$ & .10 & 1 & 1 & 1 & .92 & 1 & .96 \\ 
  .50 & $10^{-3}$ & .20 & 1 & 1 & 1 & .92 & 1 & .96 \\ 
  .50 & $10^{-3}$ & .30 & 1 & 1 & 1 & .92 & 1 & .96 \\  
  \bottomrule
\end{tabular}
\label{tab:performanceResultsBICLogLikelihood}
\end{table}

\subsection{Case Study} \label{sec:resultsRealWorldDataApplications}

This section presents the outcomes of the case study (see Section~\ref{sec:caseStudy}), where the proposed method was applied to the four drug-ADR pairs introduced above. Three of these drug-ADR pairs are known to be associated, while the combination of antibiotics and GI bleeding serves as our negative control.
The data set contains information on over 1.2 million insurants from 2004 until 2017. Given that the data are in quarter-years and we consider a time period of 14 years, there are a total of $T = 56$ time points.

We treat drug dispensation as synonymous with drug exposure. This is a strong assumption for several reasons, e.g., it assumes perfect adherence and overlooks potential changes in the treatment plan.

Table~\ref{tab:numberOfPatientsCaseStudy} displays four $2 \times 2$ contingency tables, one for each drug-ADR pair. In each table, the entries represent the number of individuals who were dispensed the drug at least once and/or experienced the ADR at least once. For example, the observed number of patients who were dispensed penicillin at any time and experienced anaphylactic shock during their coverage period is 25. Similarly, the count of patients who experienced anaphylactic shock throughout their observed period but were not dispensed penicillin is 171, and so forth.

\begin{table}[h!]
    \caption{The $2 \times 2$ contingency tables for the four drug-ADR pairs considered in the case study. Each table shows the number of patients that were dispensed the drug and/or experienced the ADR.}
    \label{tab:numberOfPatientsCaseStudy}
    \small
    \begin{subtable}{.5\linewidth}
      \centering
      
        \begin{tabular}{c c c c}
            \toprule
                &  ADR & not ADR & \textit{total}\\ 
            \midrule
            drug           & 25 & 74,068 & 74,093 \\
            not drug       & 171 & 1,179,093 & 1,179,264 \\ \midrule
            \textit{total} & 196 & 1,253,161 & 1,253,357 \\ \bottomrule
        \end{tabular}
        \caption{Penicillin and anaphylaxis}
    
    \end{subtable}%
    \begin{subtable}{.5\linewidth}
      \centering
        
        \begin{tabular}{c c c c}
            \toprule
                &  ADR & not ADR & \textit{total}\\ 
            \midrule
            drug           & 852 & 16,111 & 16,963 \\
            not drug       & 10,252 & 1,226,142 & 1,236,394 \\ \midrule
            \textit{total} & 11,104 & 1,242,253 & 1,253,357 \\ \bottomrule
        \end{tabular}
        \caption{DOACs and GI bleeding}
        
    \end{subtable} \\[.5cm] 
    
    \begin{subtable}{.5\linewidth}
      \centering

        \begin{tabular}{c c c c}
            \toprule
                &  ADR & not ADR & \textit{total}\\ 
            \midrule
            drug           & 2,079 & 59,515 & 61,594 \\
            not drug       & 6,634 & 1,185,129 & 1,191,763 \\ \midrule
            \textit{total} & 8,713 & 1,244,644 & 1,253,357 \\ \bottomrule
        \end{tabular}
        \caption{Antipsychotics and type 2 diabetes}
        
    \end{subtable}%
    \begin{subtable}{.5\linewidth}
      \centering

      \begin{tabular}{c c c c}
            \toprule
                &  ADR & not ADR & \textit{total}\\ 
            \midrule
            drug           & 9,082 & 716,990 & 726,072 \\
            not drug       & 2,022 & 525,263 & 527,285 \\ \midrule
            \textit{total} & 11,104 & 1,242,253 & 1,253,357 \\ \bottomrule
        \end{tabular}
        \caption{Antibiotics and GI bleeding}
    \end{subtable}%
\end{table}

Figure~\ref{fig:resultsCaseStudyBIC} shows the BIC scores for all eight exposure models for the four drug-ADR pairs. The models are arranged from the best (based on the BIC) on the left to the worst on the right. In all figures, except for the lower right plot, the BIC value for the no association model is notably higher than the other BIC scores, surpassing the limits of the $y$-axis. We have included the rounded value of the BIC in white on the corresponding bars. We discuss each drug-ADR pair individually.

\begin{figure}[h!]
    \centering
    \includegraphics[width=\textwidth]{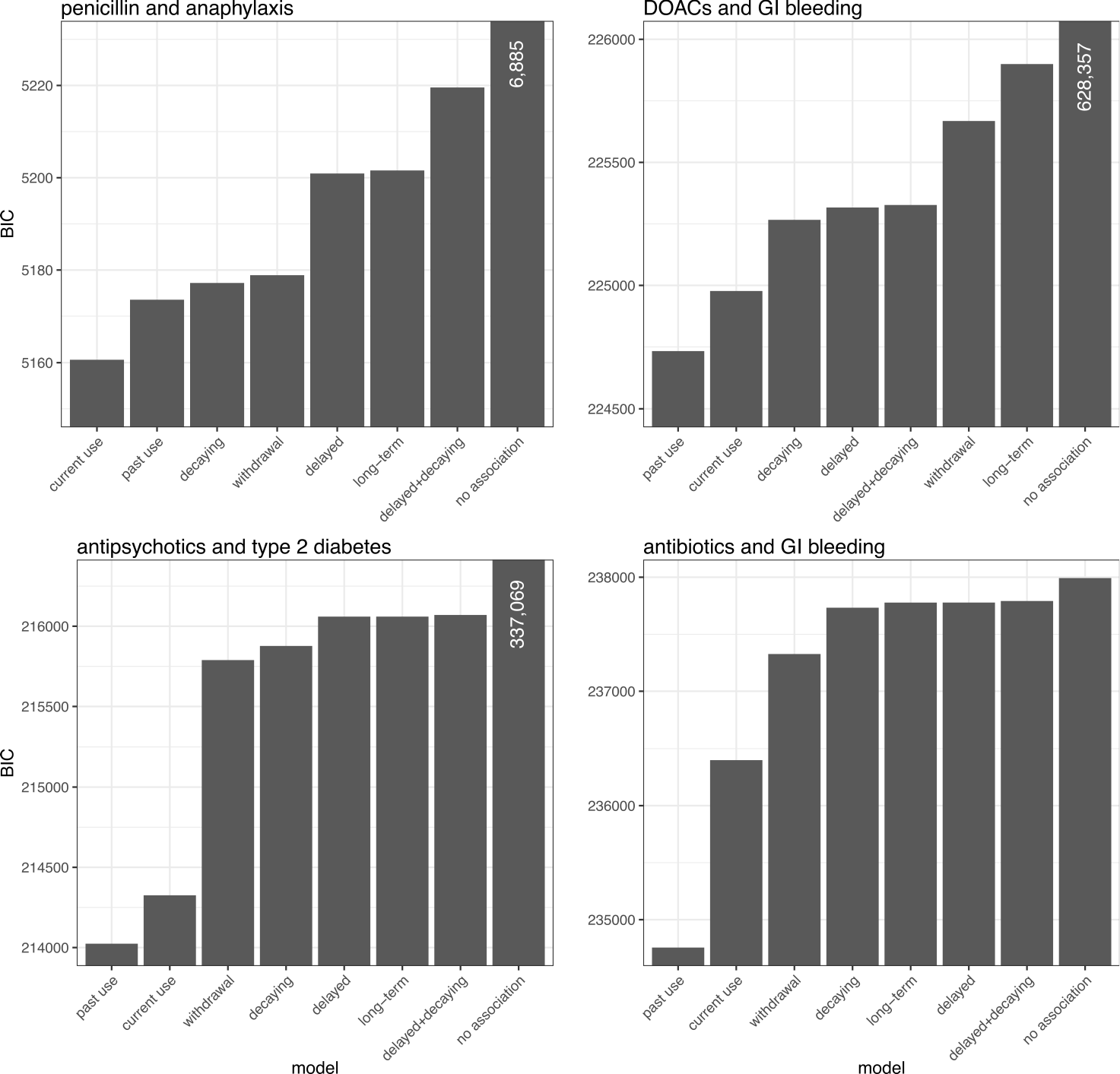}
    \caption{The BIC values for the eight exposure models outlined in Section~\ref{sec:exampleExposureModels} across the four drug-ADR pairs examined in the case study. The models are arranged from the best (based on the BIC) on the left to the worst on the right.
    In three of the four cases, the BIC value for the null model far exceeds the range of $y$-axis. The rounded BIC values are added to the respective bars.}
    \label{fig:resultsCaseStudyBIC}
\end{figure}

The results for penicillin and anaphylaxis, presented in the upper left corner, align with the anticipated model: current use. The BIC score, and consequently the posterior probability of the null model, strongly indicate the presence of an association. As mentioned earlier, although the decaying model may be more appropriate, the time resolution in quarter years is inadequate for distinguishing between this model and the current use model.

For DOACs and GI bleeding, the past use model attains the best BIC score. Analogous to the previous drug-ADR pair, the BIC score for the null model substantially exceeds the scores of the other models, suggesting a strong association.

A similar pattern emerges for antipsychotics and type 2 diabetes, with the past use model yielding the best fit, closely followed by the current use model. Once more, the inadequacy of the null model's fit implies an association between the drug and ADR.

For the negative control, antibiotics and GI bleeding, the past use model performs best as well. Despite the null model having the least favorable performance, its BIC score is comparable to the scores of the other models. In contrast, for the other three drug-ADR pairs, the difference between the null model's BIC score and the BIC scores of the other models was much larger. 

To investigate why the past use model is clearly preferred in three out of the four drug-ADR pairs~ under consideration, we delve deeper into this preference and examine the BIC values for the past use exposure model across all values of the parameter $p$. This exposure model represents a scenario where the patient is at maximal risk during and for an extended period after dispensation, where the length of the period equals $p$. See equation~\eqref{eq:modelPast} for the definition and Figures \ref{fig:riskModels}k and \ref{fig:riskModels}l for examples. The parameter $p$ ranges from $1$ to $T-1$ (equaling $55$ in our case).
Figure~\ref{fig:resultsCaseStudyPastParameters} presents the BIC values for both the current use model and past use models across all permissible values of $p$. The current use model is represented in orange and positioned at $p = 0$ since the past use model is, in that case, equivalent to the current use model. 

\begin{figure}[h!]
    \centering
    \includegraphics[width=\textwidth]{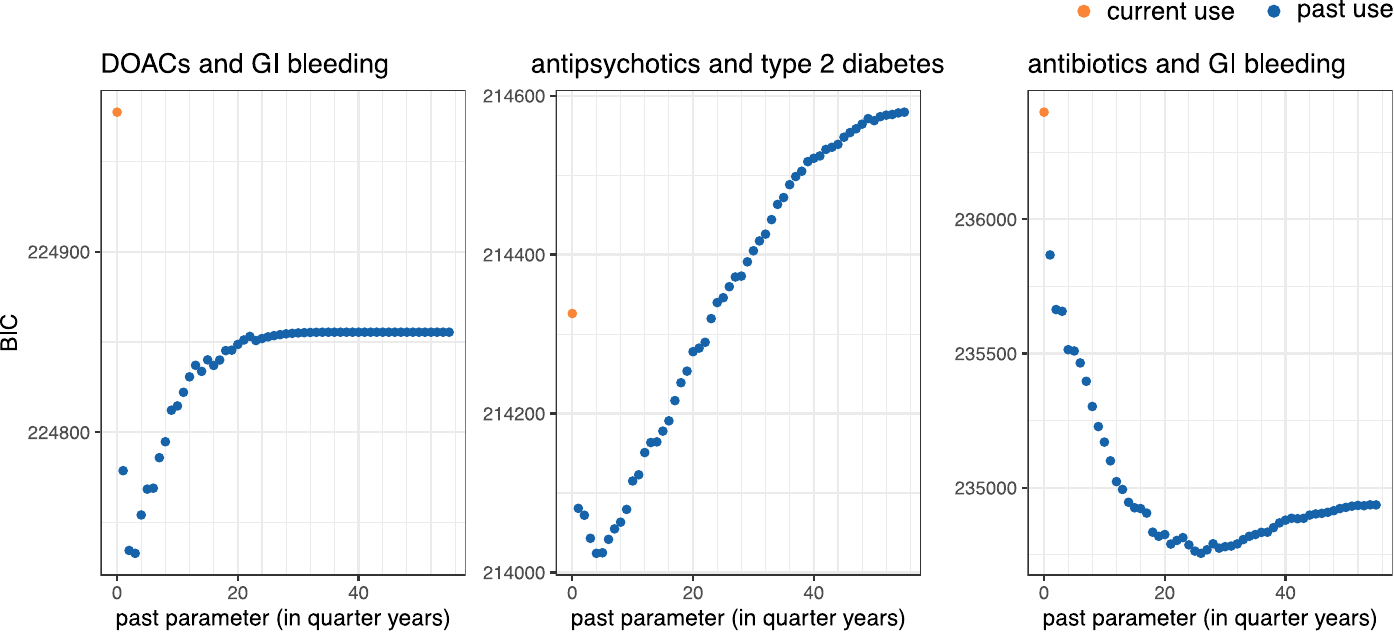}
    \caption{The BIC values for the current use and past use models for three drug-ADR pairs, for all possible parameter values of $p$ (corresponding to quarter years). The value of the current use model is depicted in orange and positioned at $p = 0$, as the past use model is equivalent to the current use model for that specific parameter value.}
    \label{fig:resultsCaseStudyPastParameters}
\end{figure}

For DOACs and GI bleeding, the BIC score reaches its minimum at $p = 3$, suggesting that the risk of experiencing GI bleeding remains elevated for approximately 3 quarters after the last dispensation. A similar trend is noted for antipsychotics and type 2 diabetes, with the lowest BIC value occurring at $p = 4$, corresponding to one year. In other words, the risk of being diagnosed with type 2 diabetes stays heightened and decreases after one year following the last dispensation.

The change in BIC scores with the parameter $p$ for the negative control, antibiotics and GI bleeding, reveals an interesting pattern that may partially explain why the past use model is favored in a scenario where the drug and the ADR considered are unrelated. The minimum occurs around $p = 26$, corresponding to 6 or 7 years after the last dispensation. It is highly improbable that exposure to antibiotics increases the risk of GI bleeding so many years later. The method likely detects the natural increase in risk with age, rather than indicating a genuine relationship between antibiotics and the ADR GI bleeding.

\section{Conclusions and Discussion}\label{sec:conclusions}

In this paper, we introduced a versatile exposure model framework designed to capture various longitudinal relationships  that can occur between a drug and an ADR. The framework allows for the estimation of model parameters based on EHC data using maximum likelihood. We suggest the utilization of the BIC to select the most suitable model. Furthermore, the BIC has a direct connection with the models' posterior probabilities, see Section~\ref{sec:modelSelection}. 
This feature makes the approach applicable in a pharmacovigilance context, where the posterior probability of the null model, representing no association, can be used to decide whether a drug-ADR pair yields a signal. Additionally, it facilitates the use of Bayesian false discovery rate procedures \citep{storey2003} to create a shortlist for the committee of medical experts. A main feature of the proposed framework is that it cannot only aid in signal detection, but also allows for exploring the nature of the relationship between the drug and ADR. 

We explored the effectiveness of this approach through a simulation and case study. For the simulation study, we developed a unique EHC data simulator capable of simulating any exposure model.

The simulation study demonstrates the capacity to, under certain conditions, determine the presence of an association between a drug and ADR, and accurately identify the correct exposure models. Key factors influencing the performance include: 1) the number of patients that were exposed to the drug ($\mu_E$), 2) the probability of experiencing the ADR when the patient is at maximal risk ($\pi_1$), and 3) the disparity between the probabilities of experiencing the ADR when the patient is at maximal and minimal risk ($|\pi_1 - \pi_0|$). Detecting signals becomes challenging when both the number of exposed patients (approximately 1\%) and the frequency with which the ADR occurs are low. Performance improves with a higher number of exposed patients and/or increased ADR frequency. 

The performance in confirming an association and identifying the correct model improves when there is a larger difference between the probabilities of the ADR occurring when at minimal and maximal risk. However, this principle does not hold for the withdrawal model. Exposure models reflecting delayed effects are frequently misidentified as the past use model, see Section~\ref{sec:resultsSimulationStudy}. This misattribution can occur since, for such models, the ADR risk increases when the exposure starts and remains high until after exposure, not unlike the trend modeled by the past use model (see Figure~\ref{fig:riskModels}). Correctly identifying the long-term model poses a significant challenge under almost all of the considered simulation settings. For an interactive exploration of all simulation study results, visit \url{https://exposuremodels.bips.eu}.

In the case study outlined in Sections~\ref{sec:caseStudy} and \ref{sec:resultsRealWorldDataApplications}, we applied the exposure model method to four drug-ADR pairs where the true temporal relationships are, at least approximately, known. Utilizing a data set consisting of insurants from two German SHIs, totaling over 1.2 million individuals, we successfully identified penicillin and anaphylaxis, where the current use model is chosen based on the BIC/posterior probability. For the other three drug-ADR pairs, the past use model was selected. While this is partly expected for the pairs DOACs and GI bleeding, and antipsychotics and type 2 diabetes, it was surprising for the negative control, antibiotics and GI bleeding.

When examining the BIC scores associated with the parameter values of the past use model, see Figure~\ref{fig:resultsCaseStudyPastParameters}, we found that for the drug-ADR pairs DOACs and GI bleeding, as well as antipsychotics and type 2 diabetes, the optimal value tends to center around one year after the last dispensation. In contrast, for the negative control, the optimal value lies around 6 to 7 years. The preference for the past use model in the negative control case appears to be driven by the natural increase in the risk of GI bleeding with age, rather than indicating a genuine relationship between the drug and the ADR. The assumption of the exposure model that the baseline risk ($\pi_0$) remains constant over time may lead to misclassification. It would be interesting to explore how the age of an individual could be included to address this issue. 

Prior efforts to employ exposure models were undertaken by Van Gaalen et al. \citep{vangaalen2015,vangaalen2017}. However, their approach considers a limited number of exposure models, and the process of estimating model parameters based on the available data is not clearly defined. Furthermore, the application of their method in a pharmacovigilance context is unclear as well.

One aspect to consider in the current study is that it does not include a comparison with other signal detection methods available in the literature\citep{dijkstra2022discovery}. To undertake such a comparison, a significant expansion of the simulation set-up is required. This expansion would involve simulating multiple drug-ADR pairs with varying exposure models simultaneously. Moreover, it is important to take into account different types of thresholds used to define a signal, as these can vary considerably from one method to another \citep{deshpande2010}. We plan to explore this comparison in future research.

The formalization of EHC data, see Section~\ref{sec:definitionEHCData}, disregards differences in dosage \citep{vangaalen2015} and only captures whether the patient was exposed or not. One could account for dosages by defining the drug history as a real-valued random vector, i.e., $\bm{X}^k \in \mathds{R}^{T_k}_+$, rather than a binary one. The reason why we opt for a binary representation is that other signal detection methods do not account for dosage as well, with notable exception of the work by Van Gaalen et al.\citep{vangaalen2015}.

As discussed in Section~\ref{sec:definitionExposureModel}, we assume that the occurrences of an ADR at different time points are independent given the drug exposure history. However, this assumption may be particularly strong for certain types of ADRs, such as anaphylaxis and myocardial infarction, where a patient is unlikely to be treated with the same drug again after experiencing such a  reaction. One potential extension to the model is to incorporate not only the drug history but also the ADR history. This extension involves modeling the conditional probability distribution $\p(\bm{Y}(t) \mid \bm{X}(1:t), \bm{Y}(1:t-1))$ rather than just $\p(\bm{Y}(t) \mid \bm{X}(1:t))$. While our framework could accommodate this extension, it would drastically increase its complexity. 
Nevertheless, it could prove to be a valuable avenue for future research.

Furthermore, it would be intriguing to explore the influence of model misspecification, particularly in scenarios where the true simulated model deviates from the predefined set of models considered by the method. Investigating such scenarios can provide insights into the method's ability to detect associations between a drug and ADR, even when temporal relationships differ from those explicitly considered. Encouragingly, the results of the case study suggest that, to some extent, the correct exposure model can still be identified even in the presence of disparities between the true and selectable models.

In theory, there is no restriction on the number of exposure models that can be simultaneously considered. Nevertheless, it is important to take into account that as the number of exposure models increases, so does the likelihood of selecting one of them over the null model. As a result, the likelihood of generating a signal for a drug and ADR increases with the number of exposure models. One potential approach to address this is by applying a false discovery rate correction to the models for each drug-ADR pair individually. However, the challenge lies in determining how to incorporate this correction alongside a false discovery rate control procedure for all drug-ADR pairs when creating a shortlist.

Employing a Bayesian approach for estimating exposure model parameters provides the advantage of incorporating prior knowledge into the modeling process. This is particularly valuable when existing knowledge is available, e.g., the probabilities of experiencing the ADR at maximal and minimal risk are close to zero. Similarly, applying a Bayesian prior to the exposure models themselves enables consideration of how frequently a specific temporal relationship is expected to occur. However, it is challenging to select an appropriate prior, given the difference in power to detect various models, as seen in the simulation study, see Section~\ref{sec:resultsSimulationStudy}. 

Exploring alternative methods for model selection beyond the BIC could be worthwhile. Even though the BIC has the advantage of being related to the posterior probability, it might be beneficial to consider other selection techniques, particularly since the BIC appears to be excessively conservative.

\subsection*{Author contributions}

The contributions are organized according to the Contributor Roles Taxonomy (CRediT), see \url{https://credit.niso.org/}. \\

\noindent \textbf{LD:} conceptualization, data curation, formal analysis, investigation, methodology, software, visualization, writing -- original draft, writing -- review \& editing. \textbf{TS:} conceptualization, resources, investigation, validation, writing -- review \& editing. \textbf{RF:} conceptualization, data curation, funding acquisition, investigation, project administration, resources, supervision, writing -- original draft, writing -- review \& editing. 

\subsection*{Acknowledgments}
We gratefully acknowledge Oliver Scholle for his insightful contributions to the application study and the interpretation of the results. 
The authors would also like to thank the statutory health insurances, namely hkk Krankenkasse and AOK Bremen/Bremerhaven, for providing the data used in this study. 

\subsection*{Financial disclosure}

This work was supported by the innovation fund (`Innovationsfonds') of the Federal Joint Committee in Germany (grant number: 01VSF16020).

\section*{Conflict of interest}

LD, TS and RF are working at an independent, non-profit research institute, the Leibniz Institute for Prevention Research and Epidemiology -- BIPS. Unrelated to this study, BIPS occasionally conducts studies financed by the pharmaceutical industry. Almost exclusively, these are post-approval safety studies (PASS) requested by health authorities. The design and conduct of these studies as well as the interpretation and publication are not influenced by the pharmaceutical industry. The study presented was not funded by the pharmaceutical industry and was performed in line with the ENCePP Code of Conduct.

\bibliographystyle{abbrvnat}
\bibliography{main}





\end{document}